\documentclass[%
reprint,
superscriptaddress,
 amsmath,amssymb,
 aps,
]{revtex4-2}

\usepackage{graphicx}
\usepackage{hyperref}
\usepackage{dcolumn}
\usepackage{bm}
\usepackage{color}
\usepackage{braket}
\usepackage{blindtext}
\usepackage{ragged2e} 
\usepackage{lipsum}

\DeclareUnicodeCharacter{2212}{-}

\begin{document}

\preprint{APS/123-QED}

\title{Quantum Hall Antidot as a Fractional Coulombmeter}

\author{Mario Di Luca}
\author{Emily Hajigeorgiou}
\author{Zekang Zhou}
\affiliation{Institute of Physics, École Polytechnique Fédérale de Lausanne (EPFL), CH-1015 Lausanne, Switzerland}

\author{Tevž Lotrič}
\affiliation{Rudolf Peierls Centre for Theoretical Physics, Parks Road, Oxford, OX1 3PU, UK}

\author{Tengyan Feng}
\affiliation{Institute of Physics, École Polytechnique Fédérale de Lausanne (EPFL), CH-1015 Lausanne, Switzerland}

\author{Kenji Watanabe}
\affiliation{Research Center for Functional Materials, National Institute for Materials Science, 1-1 Namiki, Tsukuba 305-0044, Japan}

\author{Takashi Taniguchi}
\affiliation{International Center for Materials Nanoarchitectonics,
National Institute for Materials Science, 1-1 Namiki, Tsukuba 305-0044, Japan}

\author{Steven H. Simon}
\affiliation{Rudolf Peierls Centre for Theoretical Physics, Parks Road, Oxford, OX1 3PU, UK}

\author{Mitali Banerjee}
\email{mitali.banerjee@epfl.ch}
\affiliation{Institute of Physics, École Polytechnique Fédérale de Lausanne (EPFL), CH-1015 Lausanne, Switzerland}
\affiliation{Center for Quantum Science and Engineering (QSE Center), École Polytechnique Fédérale de Lausanne (EPFL), CH-1015 Lausanne, Switzerland}

\begin{abstract}

The detection of fractionally charged quasiparticles, which arise in the fractional quantum Hall regime, is of fundamental importance for probing their exotic quantum properties. While electronic interferometers have been central to probe their statistical properties, their interpretation is often complicated by bulk–edge interactions. Antidots, potential hills in the quantum Hall regime, are particularly valuable in this context, as they overcome the geometric limitations of conventional designs and act as controlled impurities within a quantum point contact. Furthermore, antidots allow for quasiparticle charge detection through straightforward conductance measurements, replacing the need for more demanding techniques. In this work, we employ a gate-defined bilayer graphene antidot operating in the Coulomb-dominated regime to study quasiparticle tunneling in both integer and fractional quantum Hall states. We show that the gate-voltage period and the oscillation slope directly reveal the charge of the tunneling quasiparticles, providing a practical method to measure fractional charge in graphene. We report direct measurements of fractional charge, finding $q = e/3$ at $\nu = 4/3$, 5/3 and 7/3, $q = 2e/3$ at $\nu = 2/3$ and $q = 3e/5$ at $\nu = 3/5$, while at $\nu = 8/3$ we observe signatures of both $e/3$ and $2e/3$ tunneling charge. The simplicity and tunability of this design open a pathway to extend antidot-based charge measurements to other van der Waals materials, establishing antidots as a powerful and broadly applicable platform to study the quantum Hall effect.

\end{abstract}

\maketitle

The fractional quantum Hall effect (FQHE) provides a versatile platform for exploring topological states of matter, characterized by the presence of fractionally charged quasiparticles. These excitations, known as anyons, are predicted to show different quantum statistical behavior than bosons and fermions. The direct way to substantiate their statistical behavior is through braiding experiments using electronic interferometers \cite{Ji2003Mar, Neder2007Jul, McClure2009Nov, Ofek2010Mar, Wei2017Aug, Deprez2021May, Ronen2021May, Fu2023Jan} where the braiding of the Abelian anyons should be manifested as phase jumps in the Aharonov-Bohm oscillations. The first such result was reported in 2019 for $\nu = 1/3$ in a GaAs heterostructure \cite{Nakamura2019Jun, Nakamura2020Sep} in a Fabry-Perot interferometer (FPI) and recently in graphene-based FPIs \cite{Werkmeister2025Apr, Samuelson2024Mar, Kim2024Nov}. While phase jumps have been consistently reported, their interpretation comes with two key caveats: (i) the jumps occurred at random intervals rather than being controlled by the number of quasiparticles in the interferometer cavity; (ii) no simultaneous confirmation of the quasiparticle charge was established. More recently, experiments in a chiral Mach-Zehnder interferometer (cMZI) in GaAs heterostructures in FQHE, incorporating an antidot to control the number of bulk quasiparticles, have demonstrated that phase jumps occur concurrently as the number of quasiparticles changes in the bulk \cite{Ghosh2025Jul, Ghosh2025Jun}. This work uses shot noise measurements to further confirm the quasiparticle charge involved in the braiding. While the results of the cMZI are significant for particle-like FQHE states (e.g. $\nu = 1/3$), the presence of bulk–edge interactions, as in conventional FPIs, obscures the outcomes in hole-like FQHE states (e.g. $\nu = 2/3$). In particular, experiments have observed quasiparticle bunching and antibunching in states hosting upstream neutral modes \cite{Ghosh2025Jun, Kim2024Dec}. These complications place significant constraints on the use of either interferometer geometry for probing the quantum properties of non-Abelian quasiparticles in exotic states such as $\nu = 5/2$ \cite{Banerjee2018Jul}.

\begin{figure*}[htb!]
 \includegraphics[width = \textwidth]{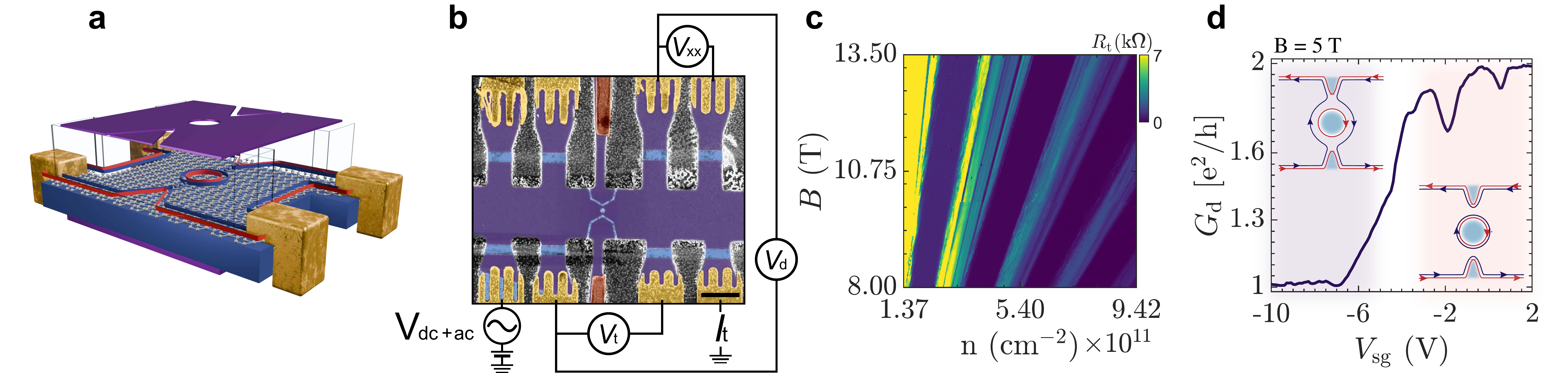}
 \centering
 \caption{\textbf{Gate-defined AD in bilayer graphene.} \textbf{a}, Schematic of the device at $\nu = 2$ showing interaction between the inner extended edge and AD-bound states (blue), while the outer edge state (red) is fully transmitted. The graphite gates are shown in purple and the bottom hBN in blue. Only the edges of the top hBN are shown for clarity. \textbf{b}, False-color scanning electron microscope image of an AD device. The ohmic contacts are shown in gold, the palladium contacts to the graphite top gate in dark orange, the graphite in purple, and in blue the top hBN is visible in places where the top graphite has been etched. The AD area is defined by the etched circle in the top graphite. A voltage of V$_\mathrm{ac} = 10$ $\mu$V (3 $\mu$V) was applied for the integer (fractional) oscillations, with measurement of the transmitted ($I_\mathrm{t}$) current. Voltage was measured across the AD ($V_\mathrm{t}$), and diagonally ($V_\mathrm{d}$), as well as on the side of it ($V_\mathrm{xx}$). The scale bar is 2 $\mu$m. \textbf{c}, Transmitted resistance $R_\mathrm{t}$ fan diagram measured across the AD as a function of the carrier density $n$ and the magnetic field $B$. \textbf{d}, Diagonal conductance for $\nu = 2$ as a function of the side gate voltage $V_\mathrm{sg}$. Each inset schematically represents the system at the corresponding shaded voltage range.}
 \label{fig:device}
\end{figure*}

Beyond the observation of interference, controlled manipulation of individual anyons, as well as verification of their fractional charge $e^*$ is needed. Direct detection of these charges is therefore a crucial step toward establishing their exotic quantum properties. In a quantum Hall antidot (AD) anyons are localized around a potential hill \cite{LevySchreier2016Aug, Sim2008Feb, Ihnatsenka2009Sep}. ADs have been extensively studied in GaAs heterostructures in the integer quantum Hall regime \cite{Hwang1991Dec, Ford1994Jun, Kataoka1999Jul, Sim2003Dec}, and they have played a crucial role in the first detection of fractionally charged quasiparticles \cite{Goldman1995Feb, Goldman2005Apr, Kou2012Jun}. Moreover, ADs are predicted to be an ideal platform for studying non-Abelian states \cite{DasSarma2005Apr, Nayak2008Sep} and controllably braid non-Abelian quasiparticles \cite{Simon2000Jun, Kivelson2024Mar}.

Despite its simple geometry, not much has been explored in graphene due to the challenges of fabricating high-quality tunable ADs in van der Waals heterostructures. The few efforts in monolayer graphene have been limited to the study of localized edges via scanning tunneling microscopy \cite{Jung2011Mar, Gutierrez2018Aug, Walkup2020Jan} or scanning gate microscopy \cite{Moreau2021Jul} and lithographically etched structures in hBN-encapsulated or suspended graphene \cite{Mills2019Dec, Mills2020Nov}. However, these approaches lack the tunability to control the coupling between the extended edges and the AD-bound states. 

Here, we present a dual gate-defined bilayer graphene AD, where the AD is fully electrostatically defined and controlled by top, bottom, and two side graphite gates, providing tunability beyond previous designs. The AD is operated in the Coulomb interaction-dominated regime as a platform for measuring the charge of the quasiparticle tunneling through the fractional bulk. We observe Coulomb-dominated oscillations at both integer and fractional filling factors, and extract the quasiparticle charge directly from the conductance oscillations as a function of carrier density and magnetic field. Our results are in good agreement with earlier studies done in GaAs heterostructures for the lowest Landau levels \cite{Kou2012Jun}. The key advantage of this geometry lies in its simplicity, making it adaptable to other systems, such as rhombohedral multilayer graphene, for probing quasiparticle charge in fractional Chern insulators \cite{Choi2025Mar, Xie2025Jul}. 

\section*{Device description}

An antidot is a potential hill introduced into a two-dimensional electron gas in a perpendicular magnetic field. In the quantum Hall regime, chiral edge modes encircle the AD while extended channels carry current along the device boundaries. The main effect of the AD potential is to lift the degeneracy of the Landau levels (LLs) encircling it, by splitting the energy spectrum into quantized levels, where each can host one charge carrier \cite{Sim2008Feb, Gutierrez2018Aug, Walkup2020Jan}. The energy splitting is related to the velocity of the edge mode $v$ and the diameter of the AD $D_\mathrm{AD}$, $\delta \epsilon \sim \hbar v/D_\mathrm{AD}$ \cite{Mills2019Dec}. By tuning an external parameter, such as the magnetic field, the top or bottom gate voltages, it is possible to control the number of quasiparticles confined in the AD-bound states and to detect them through resonant tunneling between the extended edge channels and the quantized AD-bound states. 

\begin{figure*}[hbt!]
 \includegraphics[width = \textwidth]{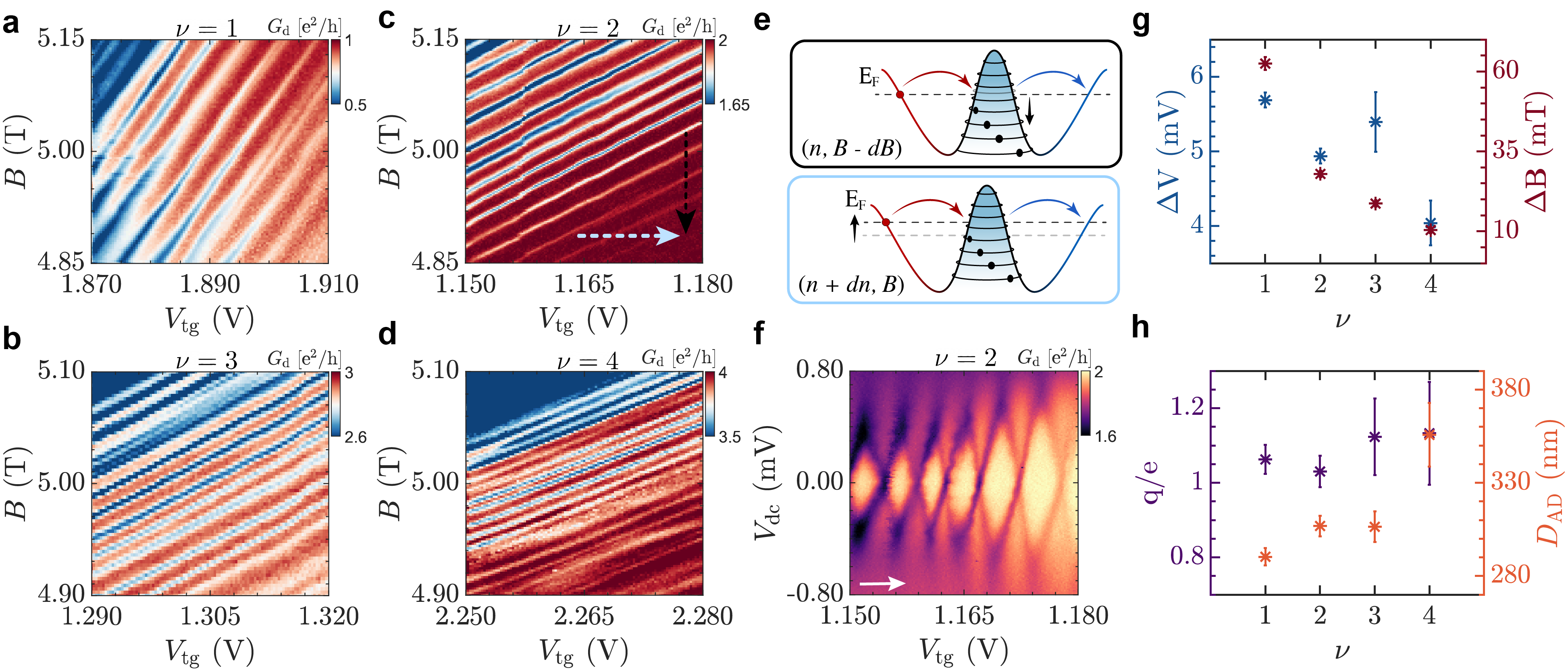}
 \begin{center}
 \caption{\textbf{Coulomb dominated oscillations for integer filling factors.} \textbf{a-d}, Diagonal conductance $G_\mathrm{d}$ oscillations as a function of the top gate voltage ($V_\mathrm{tg}$) and the magnetic field ($B$) for the inner-edge of filling factors $\nu = 1-4$. \textbf{e}, Schematic of the oscillation mechanism when varying the magnetic field (top panel) or the charge carrier density (bottom panel), corresponding respectively to the black and light blue arrows in panel \textbf{c}. \textbf{f}, Diagonal conductance $G_\mathrm{d}$ as a function of top gate voltage and dc bias displaying characteristic Coulomb diamonds for $\nu = 2$ at B = 5 T. The white arrow in panel \textbf{f} indicates the direction of increasing number of electrons bound to the AD. \textbf{g}, Top gate period $\Delta \mathrm{V}$ (blue) and magnetic field period $\Delta \mathrm{B}$ (red) as a function of the filling factor $\nu$. \textbf{h}, Measured tunneling charge (purple) and AD diameter (orange) as a function of the filling factor $\nu$.} 
 \label{fig:ff2}
 \end{center}
\end{figure*}

In Fig. \ref{fig:device}a, we present our gate-defined bilayer graphene (BLG) AD device. The AD is electrostatically defined by an etched hole in the top graphite gate, with its density tuned near the highly resistive $\nu = 0$ state via a negative voltage applied to the bottom graphite gate (BG). The BLG layer is encapsulated between two layers of hexagonal boron nitride (hBN), as shown in the schematic in Fig. \ref{fig:device}a. The device is fabricated using a standard van der Waals dry pick-up technique, described in the Methods section. The charge carrier density of BLG is tuned via voltages applied to both the bottom and the top graphite gates. Lithographically etched line cuts in the top graphite define the top gate (TG) and two side gates (SG) as shown in Fig. \ref{fig:device}b. The TG sets the filling factor in the bulk $\nu$, and by applying a negative voltage to the SG we control the number of transmitted edges through the constriction, defined as the ensemble of the side gates and the AD.

The devices have ohmic contacts along the path of both upstream and downstream extended edges, allowing for current injection and differential chemical potential measurements across the AD. Unless otherwise stated, all measurements were performed with an applied voltage of V$_\mathrm{ac} = 10$ $\mu$V (3 $\mu$V) for the IQHE (FQHE), while measuring the transmitted current, $I_\mathrm{t}$. Furthermore, longitudinal resistance $R_\mathrm{xx} = V_\mathrm{xx}/I_\mathrm{t}$ is measured outside the AD and the transmitted resistance across it $R_\mathrm{t} = V_\mathrm{t}/I_\mathrm{t}$, while the measurement of the diagonal voltage drop gives the diagonal conductance through the AD, $G_\mathrm{d} = I_\mathrm{t}/(V_\mathrm{d}) \equiv \nu_\mathrm{c} e^2/h$, where $\nu_\mathrm{c}$ is the constriction filling factor.

To ensure that ballistic edge channels encircle the AD, its diameter must be significantly larger than the magnetic length ($l_\mathrm{B}$), $D_\mathrm{AD} \gg l_\mathrm{B} \approx \frac{24 \text{ nm}}{\sqrt{B [\text{T}]}}$. In our devices, atomic force microscopy measurements show a lithographic diameter of the graphite AD of $D_\mathrm{AD\#1}$ = 190 $\pm$ 5 nm ($D_\mathrm{AD\#2}$ = 195 $\pm$ 5 nm) for the first device AD \#1 (second device AD \#2), see Supplementary Materials (Sec.\ref{sec:characterization}). However, because of the electric field propagation through the thicker top hBN layer, the effective area of the AD at the graphene surface is expected to be larger. Using a top hBN thickness of approximately 50 nm, we estimate the effective AD diameter to be around $\sim 300 $ nm. 

The devices are initially characterized by studying the development of LLs as a function of carrier density and magnetic field, as shown in Fig. \ref{fig:device}c. The high quality of the device is confirmed from the appearance of many well-developed fractional Hall states at B = 8 T. 

\section*{Coulomb oscillations at integer filling factors}

We begin by studying the first device (AD \#1) at a perpendicular magnetic field of B = 5 T at integer bulk filling factors, $\nu = 1-4$. By tuning the SG voltages, the innermost edge states can be brought into proximity with the AD-bound states. Figure \ref{fig:device}d shows the diagonal conductance $G_\mathrm{d}$ as a function of the side gates voltage $V_\mathrm{sg}$ at $\nu = 2$. As $V_\mathrm{sg}$ decreases, the inner edge can be selectively partitioned, crossing over from the full transmission ($V_\mathrm{sg} \approx 2$ V) to full reflection ($V_\mathrm{sg} \approx -8$ V).

When the constriction is set on a QH plateau, charge transport between the counter-propagating edges is suppressed, resulting in constant $G_\mathrm{d}$. However, when the inner extended edge is brought into proximity to the AD-bound states, $\nu_\mathrm{c} < \nu$, charge carriers can tunnel from the extended edges to the AD. When the chemical potential of an AD-bound state aligns with the Fermi level, a quasiparticle can tunnel resonantly through the AD, leading to a dip in the tunneling conductance $G_\mathrm{d}$. As the magnetic field is adiabatically varied by a small amount (by a fraction of the quantum flux), the AD-bound states adjust to maintain a constant enclosed flux. As the edge states move over the AD potential to adjust their size, their energy changes accordingly. This process generates periodic oscillations in $G_\mathrm{d}$ with a magnetic field period of $\Delta \mathrm{B} = \frac{\phi_0}{NA}$, where $\phi_0$ is the quantum of magnetic flux, $N$ is the number of tunneling quasiparticles per quantum flux, and $A = \pi D_\mathrm{AD}^2/4$ is the AD area \cite{Goldman2008Mar}. For oscillations in the IQHE, $N$ is equal to the filling factor of the interacting edge, $N = \nu_\mathrm{int}$.

The oscillations can also be controlled electrostatically using the TG and BG. By sweeping both gate voltages simultaneously, while maintaining a constant electron density $n$, the AD area can be changed. The oscillations observed using this method are provided in the Supp. Materials (Sec.\ref{sec:Area}). 

Another method involves varying the electron density around the AD while keeping its area constant, allowing for tunneling at periodic voltage values of the density-controlling gate. This method has the advantage of allowing the measurement of the tunneling charge, $\Delta \mathrm{V} = \frac{q}{CA}$, where $C$ is the capacitance per unit area of the swept gate, $q$ the tunneling charge and $\Delta \mathrm{V}$ the gate period \cite{Goldman1995Feb, Mills2019Dec, Mills2020Nov}. In our device, the density can be tuned either by adjusting the TG or BG voltage, while keeping the other gates constant. In the following sections, we vary the electron density using only the TG, with the BG fixed (for oscillations using the BG see the Supp. Materials Sec.\ref{sec:BG}).

For each filling factor, the SG voltages are set such that the transmission is $t \simeq 1$. Figure \ref{fig:ff2}a-d shows oscillations in the diagonal conductance as a function of the TG voltage, $V_\mathrm{tg}$, and the magnetic field for the inner edge of filling factor $\nu = 1-4$. A dip in the conductance appears each time carriers can transfer from the higher to the lower potential edge through the AD-bound states, as illustrated in the schematic in figure~\ref{fig:ff2}e. To determine the oscillation period, we perform a two-dimensional Fast Fourier Transform (2D-FFT) (reported in the Supp. Materials Sec.\ref{sec:IQH}) for each filling factor. From the main peak frequency, we extract the magnetic field period $\Delta \mathrm{B}$ and the TG period $\Delta \mathrm{V}$. Figure \ref{fig:ff2}g reports the periods for each filling factor. Notably, $\Delta \mathrm{V}$ slightly decreases with increasing filling factor due to the increasing number of edge states encircling it \cite{Gutierrez2018Aug}. Meanwhile, $\Delta \mathrm{B}$ monotonically decreases, as expected from the $\phi_0/N$ dependence \cite{Goldman2008Mar, Ihnatsenka2009Sep}. The latter can be used to compute the AD diameter $D_\mathrm{AD} = 2\sqrt{\phi_0/\pi \Delta \mathrm{B} N}$ reported in orange in Fig. \ref{fig:ff2}h. As previously mentioned, the measured diameter increases with the filling factor. Using the geometric capacitance of the TG, $C$, and substituting the area dependence from the magnetic field and gate period, the charge of the tunneling particle can be written as

\begin{equation}
q = \frac{C \phi_0 \Delta \mathrm{V} }{ N\Delta \mathrm{B}}
\label{eq:charge}
\end{equation}

Figure \ref{fig:ff2}h shows the computed charge (purple) for different filling factors. The calculated charge is 5\% higher than the expected value of one electron per period. We attribute this discrepancy to a small dependence of the AD potential on the TG voltage as reported in the Supp. Materials (Sec.\ref{sec:IQH}). 

Finally, we study the dc bias V$_\mathrm{dc}$ dependence of the oscillations. Figure \ref{fig:ff2}g shows the diagonal conductance oscillations as a function of the TG voltage and the dc bias for the inner mode of $\nu = 2$. The diagonal conductance $G_\mathrm{d}$ exhibits a Coulomb diamond behavior, characterized by trapezoidal regions where tunneling through the AD is suppressed. 

\section*{Coulomb oscillations at fractional filling factors}

\begin{figure*}[htb!]
 \includegraphics[width = \textwidth]{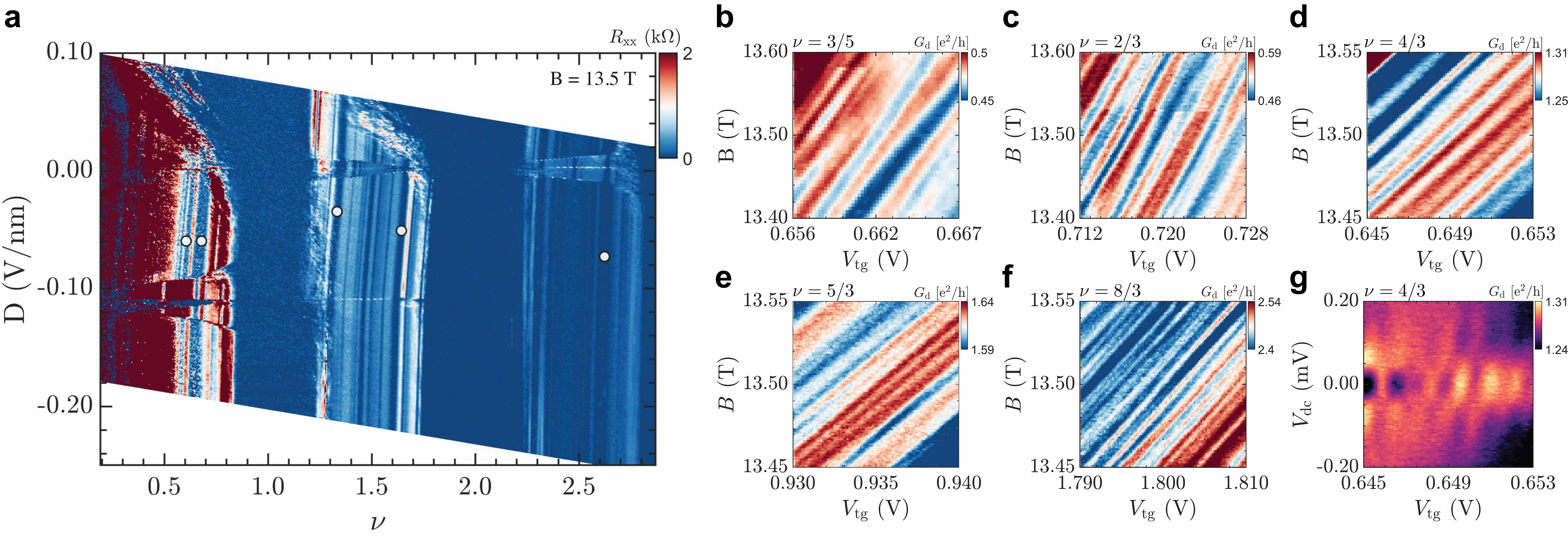}
 \begin{center}
 \caption{\textbf{Oscillations in the fractional quantum Hall regime.} \textbf{a}, Longitudinal resistance $R_\mathrm{xx}$ measured outside the AD in function of the bulk filling factor $\nu$ and the displacement field $D$ at B = 13.5 T. \textbf{b-f}, Diagonal conductance oscillations $G_\mathrm{d}$ as a function of the top gate voltage $V_\mathrm{tg}$ and the magnetic field $B$ for $\nu = 3/5, 2/3, 4/3, 5/3$ and 8/3 respectively. The white dots in panel \textbf{a} indicate the values at which the oscillations have been taken. \textbf{g}, Diagonal conductance oscillations $G_\mathrm{d}$ in function of the top gate voltage and the dc bias $V_\mathrm{dc}$ for $\nu = 4/3$.} \label{fig:fractions}
 \end{center}
\end{figure*}

Next, we turn to the FQHE regime. For this part of the study, we fabricated a second device (AD \#2) that showed a richer set of fractional quantum Hall states. Figure \ref{fig:fractions}a shows the longitudinal resistance $R_\mathrm{xx}$, measured outside the AD in function of the bulk filling factor $\nu$ and the displacement field $D$ at B = 13.5 T. This map closely resembles similar results observed in bilayer graphene, showing the emergence of strong even-denominator states, $\nu = 1/2, 3/2$ and $5/2$ in the first orbital LLs \cite{Hunt2017Oct, Li2017Oct, Li2018Jan, Huang2022Jul, Kumar2025Aug}. However, due to a highly resistive charge neutrality point, it's not possible to observe any states below $\nu = 1/2$, such as $\nu = 1/3$ and 2/5.

For each fractional filling factor, we repeat the procedure outlined for the integer case. The back gate is set to $\nu \leq 0$, and the displacement field $D$ is kept close to zero, ensuring that all states are measured under comparable conditions. Figures \ref{fig:fractions}b-f show the diagonal conductance oscillations observed at fractional filling factors $\nu = 3/5, 2/3, 4/3, 5/3$ and 8/3. The displacement field $D$ and the corresponding $\nu$ values at which these oscillations are observed are marked by white dots in Fig. \ref{fig:fractions}a. 

For all oscillations, we perform a 2D-FFT to extract the oscillation periods (see Supp. Materials Sec.\ref{sec:FQHE}) as well as the dc-bias dependence, shown in Fig. \ref{fig:fractions}g for $\nu = 4/3$. 

\begin{figure}[h!]
 \includegraphics[width = 0.5\textwidth]{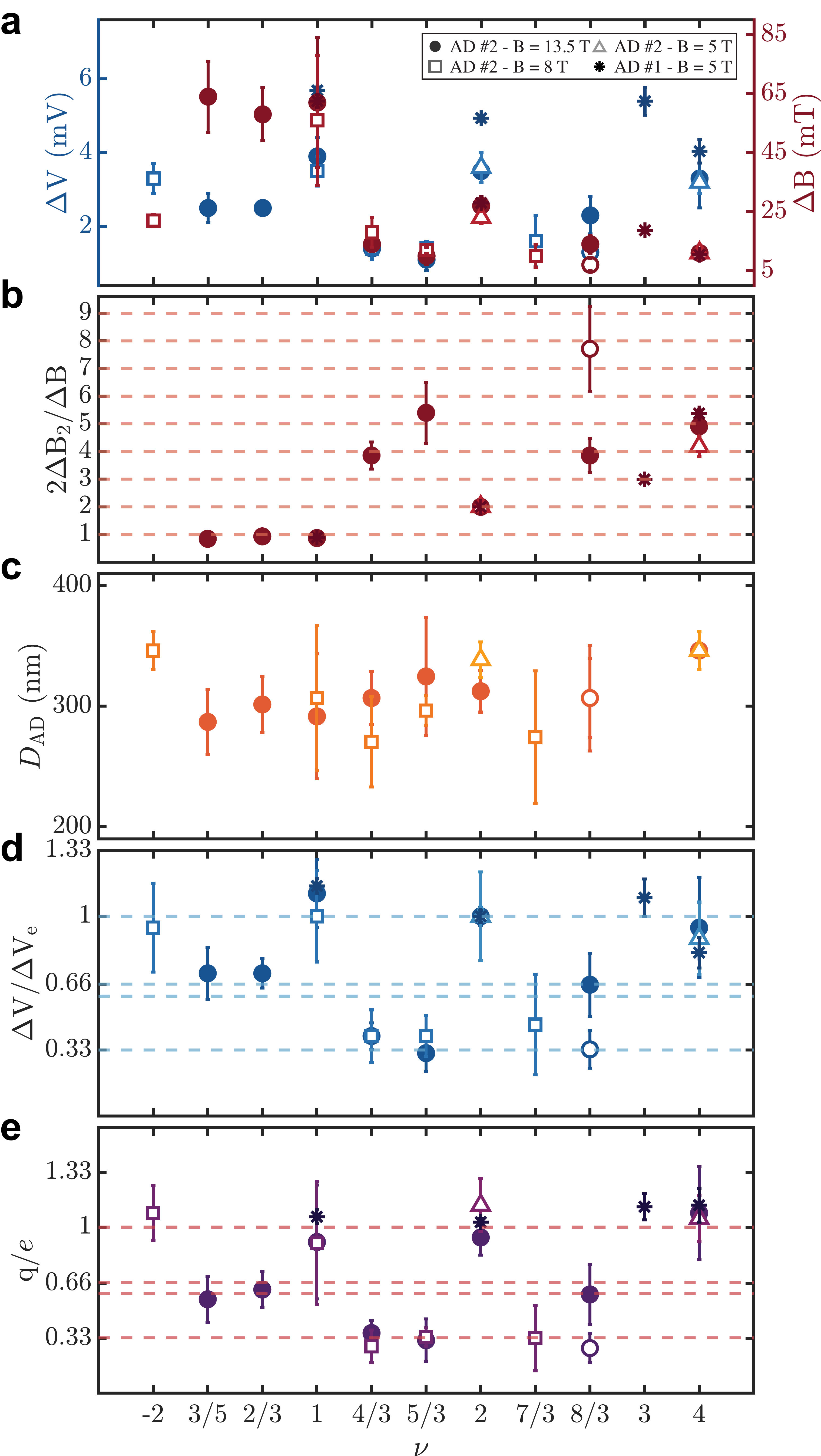}
 \begin{center}
 \caption{\textbf{Charge measurement.} \textbf{a}, Top-gate period $\Delta V$ (blue) and magnetic-field period $\Delta B$ (red) as a function of filling factor $\nu$ for oscillations measured at different magnetic fields and in two devices. \textbf{b}, Ratio of the field period at $\nu = 2$, $\Delta B_2$, to the field period at each filling factor; the factor of 2 accounts for the $\phi_0/2$ periodicity characteristic of $\nu = 2$. \textbf{c}, AD diameter as a function of filling factor $\nu$. \textbf{d}, Ratio of the gate period corresponding to a single electron, $\Delta V_e$, to the gate period at each filling factor; for AD \#2 at $B = 8$ T, $\Delta V_e = \Delta V_1$ was used, while for all other cases $\Delta V_e = \Delta V_2$. \textbf{e}, Quasiparticle charge as a function of filling factor, computed using Eq. \ref{eq:charge}. In all the panels, the empty circle indicates the smaller oscillation at $\nu = 8/3$. In panel \textbf{b}, the data for AD \#2 at B = 8 T are not reported, as measurements at $\nu =2$ were not performed; in panel \textbf{c}, data for AD \#1 are omitted for clarity as they correspond to a different device.}
 \label{fig:charge}
 \end{center}
\end{figure}

Figure \ref{fig:charge}a shows the extracted gate periods (red) and field periods (blue) for various filling factors, measured for two different devices and at different magnetic fields. For $\nu = 8/3$, measured at a magnetic field of $B = 13.5$ T, we report two data points (full and empty circles) corresponding to the two observed oscillations described in the Supp. Materials (Sec.\ref{sec:FQHE}). The smaller oscillation has almost half the period of the larger data point, corresponding to an oscillation appearing in between two larger oscillations. However, since this occurs only in certain regions, we have decided to retain both points. In the following, we refer to them as the small (empty circle) and large (full circle) $8/3$ oscillations according to their respective periods.

A decrease in the magnetic field period with the filling factor is observed, consistent with a $\phi_0/N$ dependence. The constant $N$, determined solely by the filling factor of the interacting edge, is the number of quasiparticles with charge $q$ transferred upon an increase of the magnetic flux by $\phi_0$ \cite{Goldman2008Mar}. These values can be extracted through comparison of the magnetic field periods. Assuming a constant area, the ratio of the two field periods is inversely proportional to the ratio of $N$, namely $\Delta B_\nu/\Delta B_{\nu'} = N_{\nu'}/N_{\nu}$. The comparison has been carried out using the oscillations obtained at filling factor $\nu = 2$ at their respective magnetic fields. The ratio $2\Delta B_2/\Delta B = N$ is plotted in Fig. \ref{fig:charge}b. The oscillations for AD \#2 at $B = 8$ T are not reported, as measurements at $\nu =2$ were not performed. The filling factors $\nu = 3/5, 2/3$, and 1 follow closely the $N = 1$ line, consistent with our integer results and with previous studies \cite{Kou2012Jun, Mills2020Nov}. Moreover, the filling factors $\nu = 4/3$ and $5/3$ are found close to $N = 4$ and $N = 5$, respectively. It is further observed that the large $\nu = 8/3$ oscillation falls close to $N = 4$, while the small falls close to $N = 8$, the latter agreeing with its expected value corresponding to a tunneling charge equal to the filling factor fundamental charge $q = e^* = 1/3$. For $\nu = 3$, the integer result is recovered. However, for $\nu = 4$, the observed value exceeds the expected value of $N = 4$, which can be attributed to a variation of the area, as will be discussed below.

Next, using the previously determined values of $N$, the AD diameter $D_\mathrm{AD}$ can be computed for different filling factors, as shown in Fig. \ref{fig:charge}c. Measurements from AD \#1 have been omitted for clarity, as they correspond to a different device and therefore a different diameter. All diameters are found to cluster around the same value $D_\mathrm{AD} \simeq 300$ nm, consistent with the correctly estimation of $N$. 
The ratio between the gate periods provides information on the charge of the tunneling quasiparticle. Assuming a constant area across different filling factors, the relation $\Delta V_{\nu}/\Delta V_{\nu'} = q_\nu/q_{\nu'}$ holds, where $q_\nu$ is the charge of the tunneling quasiparticle at filling factor $\nu$. Using the integer filling factors as a reference for the electron charge, this can be expressed as $\Delta V / \Delta V_{e} = q/e$, where $\Delta V_{e}$ is the gate voltage required to tunnel a single electron. Figure \ref{fig:charge}d shows this ratio for different filling factors; for AD \#2 at B = 8 T, $\Delta V_e = \Delta V_1$ was used, while for all other cases $\Delta V_e = \Delta V_2$. As expected for the IQHE, filling factors $\nu = -2, 1, 2, 3$, and $4$ consistently fall around the $q/e = 1$ line, independently of the magnetic field at which they are measured. The filling factor $\nu = 3/5$ appears close to $q/e = 3/5$, which deviates from the expected value of $q/e = 1/5$ \cite{Jain2007Mar}. Filling factors $\nu = 2/3$ and the large $8/3$ oscillation follow the $q/e = 0.66$ line, as expected for $\nu = 2/3$, and in agreement with previous results in GaAs ADs \cite{Kou2012Jun}. Finally, the filling factors $\nu = 4/3, 5/3, 7/3$ and the small $8/3$ cluster close to $q/e = 0.33$, with $\nu = 7/3$ showing a larger uncertainty due to difficulty in identifying the oscillation period. 

Finally, Eq. \ref{eq:charge} is used to compute the charge of the tunneling quasiparticles for each filling factor, employing the same values of $N$ used to determine the AD diameter. Figure \ref{fig:charge}e shows the quasiparticle charge as a function of filling factor. The values align with those obtained from the gate-period ratios in Fig. \ref{fig:charge}d, confirming that the assumption of a nearly constant area across filling factors is reasonable. For integer filling factors, $q/e$ is close to 1, as expected for tunneling of electrons. Fractional filling factors $\nu = 4/3, 5/3, 7/3$ and the small 8/3 oscillation fall near $q/e = 0.33$, while $\nu = 2/3$ and the large $8/3$ oscillation lie close to $q/e = 0.66$. Finally, $\nu = 3/5$ appears near $q/e = 3/5$.

\section*{Discussion}

Previous works on ADs have focused only on the IQHE or on the FQHE at filling factors below $\nu = 1$ \cite{Goldman1995Feb, Kataoka1999Jul, Kataoka2000Aug, Goldman2005Apr, Goldman2008Mar, Kou2012Jun, Mills2019Dec, Mills2020Nov}. In this regime, $N$ is always equal to 1, (e.g. $\nu = 2/3$) or to the filling factor of the interacting edge $N = \nu_\mathrm{int}$. However, for FQHE states in a higher LL, $N$ can take different values and should be understood as the number of quasiparticles transferred per quantum flux $\phi_0$. This interpretation is particularly clear for $\nu = 4/3$ and $\nu = 5/3$ where $N$ takes the values of 4 and 5, respectively. The edge-counting approach used in some prior works \cite{Kou2012Jun} would only imply 2 edges for $\nu = 4/3$, leading to an estimated AD diameter far larger than lithographic size between the two side gates ($\sim 450 $ nm).

Using this interpretation, we observe that for $\nu = 4/3, 5/3, 7/3$ and the small 8/3 oscillation, the measured quasiparticle charge $q/e$ is close to the expected value of the minimal excitation $e^* = 1/3$. However, for $\nu = 2/3$ and the large $\nu = 8/3$, the measured values are closer to $q/e = 2/3$. This is consistent with the only previous AD measurement at $\nu = 2/3$ in GaAs \cite{Kou2012Jun}. It is worth noting that 
shot noise measurements in GaAs, found a charge at $\nu = 2/3$ to be $q/e = 2/3$ at low temperature as well (but $q/e = 1/3$ at higher temperatures) \cite{Biswas2022Dec, Ghosh2025Jun}. However, other experiments that simultaneously measured both the auto-correlation and cross-correlation reported that, even when the auto-correlation indicated a noise charge of $q/e = 2/3$ the cross-correlation consistently showed $q/e = 1/3$, thereby ruling out quasiparticle bunching as the underlying mechanism \cite{Kapfer2018Oct}. 

As of yet, no detailed explanation has been given as to why one sometimes observes charge $q = 2/3$. However, there are some suggestive pieces of theory that lead us to an interesting interpretation. The naive picture of a 2/3 edge is a downstream charge 1 edge and an upstream charge $1/3$ edge. However, 
in a classic theoretical work by Kane, Fisher, and Polchinski \cite{KaneFisherPolchinski,KaneFisher} it was shown that in the presence of disorder, due to scattering between the edge modes, under renormalization group flow, the naive structure flows to a different structure: a single downstream charge 2/3 edge and an upstream neutral mode.
While at short length scales and high temperature, one might see the original naive structure, at the lowest temperatures and longest length scales one should only see the renormalized structure. The details of the crossover temperature depend on the strength of the disorder and the initial tunneling strength between the two edges. 

Pursuing this line of reasoning we have repeated the renormalization group calculation but starting with $n > 1$ downstream charge 1 modes and an upstream charge $1/3$ mode (details in Supp. Material Sec. \ref{sub:RG}).
Interestingly if one performs the same calculation on an edge at $\nu=n + 2/3$ for integer $n>0$, we find that under most conditions the system does not includes a charge $2/3$ mode at the lowest temperature but instead includes a $1/3$ mode. Thus at the lowest temperatures and longest length scales one should not find a 2/3 edge. However, one could also have a scenario where the $n$ integer modes are physically separated from the $2/3$ edge such that the $2/3$ renormalizes by itself first, to give the Kane-Fisher-Polchinski result over a very wide temperature range. Then only a very low temperatures the $n$ integer edges would couple and undo this restructuring of the edge to return the edge to the naive structure again. 

We argue that this scenario is potentially in agreement with our experiments as well as with prior work \cite{Kou2012Jun,Biswas2022Dec,Ghosh2025Jun}. 
While $\nu=2/3$ should always renormalize to a single charge 2/3 edge at the lowest temperatures, the observed charge at finite temperature for the $n+2/3$ edge should depend on how strongly the $n$ integer edges are coupled to the fractional $2/3$. If these are strongly coupled then one should observe charge 1/3 edge modes. However, if they are weakly coupled there is an intermediate temperature regime (which could extend well below the lowest experimental temperatures if they are very weakly coupled) where one observes a charge 2/3 edge. Turning to our experiment, the fact that the $\nu=1$ charge gap is fairly small, suggests that that edge mode is not physically well separated from the nearby fractional edges. Thus the $\nu=5/3$ edge should be in a regime where we observe a 1/3 charge mode in agreement with the observation of charge 1/3 here. However, the $\nu=2$ charge gap being large suggests that the fractional $\nu=8/3$ edge should, over a wide temperature regime, behave like a charge $2/3$ edge mode, also in agreement with 
the observation of charge 2/3. 

We emphasize that this renormalization group calculation is not meant to be a full calculation of the expected AD oscillations in this antidot device, but it does strongly point towards a direction for future theoretical analysis. 

The fact that multiple oscillation periods are observed for $\nu=8/3$ could suggest that both pieces of physics can be observed. However, we should also not neglect the possibility that coulomb charging effects, of the kind described by Rosenow and Halperin \cite{RosenowHalperin,HalperinSternNederRosenow} is playing a significant role. 

Another possible explanation for the doubling of the measured charge is that quasiparticles may exhibit different tunneling preferences for even versus odd numbers of quasiparticles in the AD. If the two oscillations have a different tunneling amplitude, and if one of them dominates over the other, we would only be able to detect the strongest one, resulting in an effective doubling of the tunneling charge.

The physics at $\nu=3/5$ is similar to that at $\nu=2/3$ in that it also renormalizes to a single charge $3/5$ mode and multiple neutral modes. Shot noise measurements have similarly shown that at low temperatures the measured Fano factor is closer to $3/5$ (similar to our measured tunneling charge), while at higher temperatures it approaches the expected value of $e/5$ \cite{Banerjee2018Jul}.

\begin{figure}[htb!]
 \includegraphics[width = 0.45\textwidth]{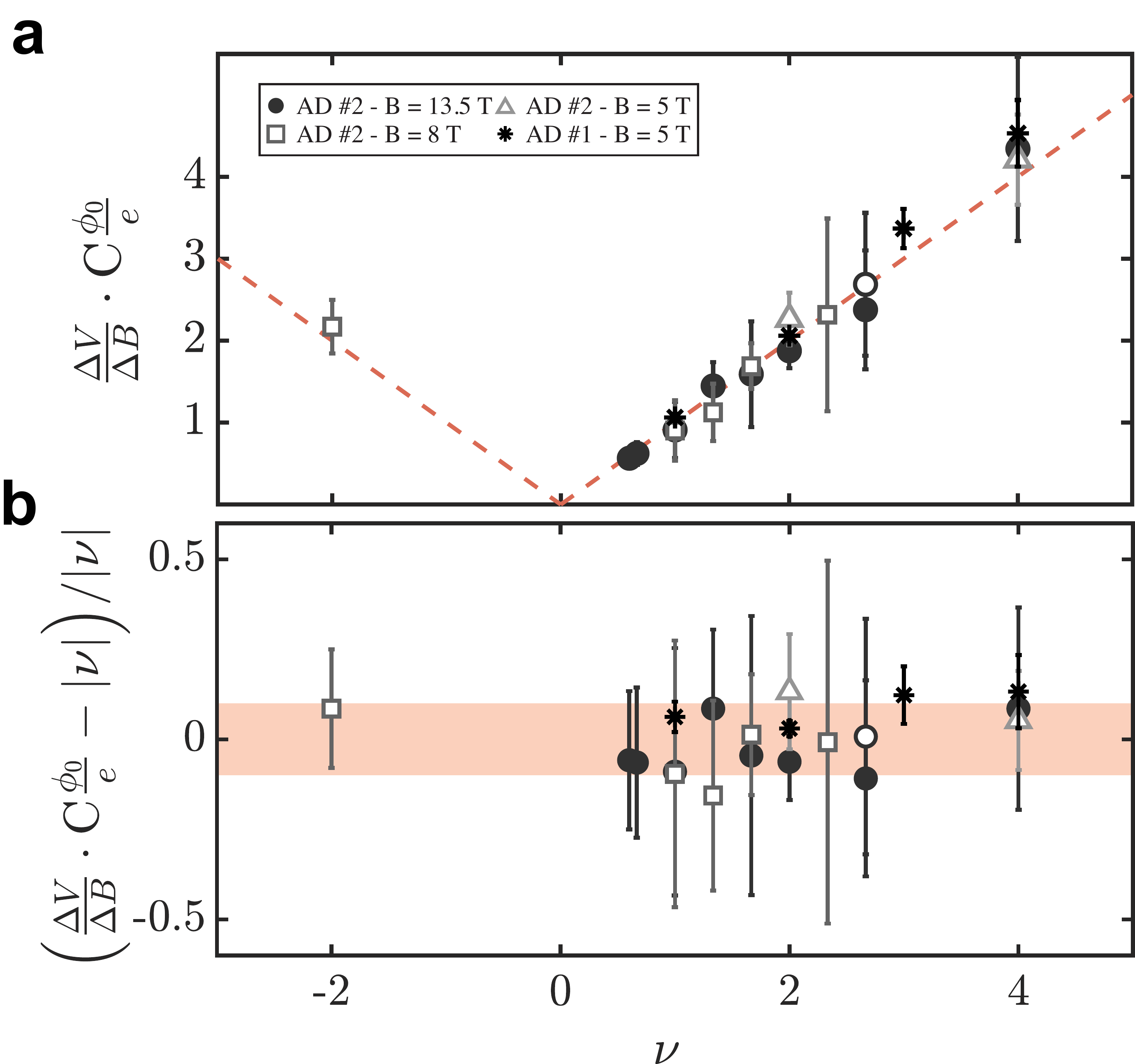}
 \begin{center}
 \caption{\textbf{Oscillation slope.} \textbf{a}, Ratio of the top-gate period to the magnetic-field period, normalized by the top-gate capacitance per unit area and $\phi_0/e$, as a function of the filling factor $\nu$. The dotted orange lines are guides to the eye. The oscillations exhibit a slope corresponding to the filling factor $\nu$. \textbf{b}, Relative error of the ratio of the top-gate period to the magnetic-field period minus the filling factor. The shaded orange band marks the $\pm 10\%$ interval.} 
 \label{fig:slope}
 \end{center}
\end{figure}

Next, using the definition of the filling factor $\nu = n\phi_0/eB$ and considering small variations $\Delta n$ (induced by $\Delta V$) and $\Delta B$, we can study the slope of the oscillations as a function of the filling factor $\nu$, as shown in Fig. \ref{fig:slope}a. The data points closely follow the orange lines, which serve as a guide to the eye, implying that the slope of the oscillations, $\Delta V /\Delta B$, always follows the filling factor at which they are measured. The fact that the oscillation slope is fixed by the filling factor implies that, once $N$ is determined, the ratio $\Delta V /\Delta B$ will always yield the fundamental charge $e^*$, since

\begin{equation}
\frac{q}{e} = \frac{C \phi_0 \Delta V}{eN \Delta B } = \frac{\nu}{N}
\end{equation}

This may lead to a misinterpretation of the calculated value, with it being ascribed to the tunneling quasiparticle charge rather than to the slope of the oscillations determined by the filling factor. This is why it is essential to compare the value obtained from the charge equation with the ratio $\Delta V /\Delta V_e$, which provides a cross-check for the correct charge determination. In our case, Figure \ref{fig:charge}d-e shows agreement among all measured filling factors, supporting the validity of this approach.

Finally, since the plotted ratio accounts only for the density variation ($\Delta n = C\Delta V$), any deviation from the orange line must arise from a change in the enclosed area, which has been assumed to be constant so far. The relative error of the slope with respect to the filling factor is shown in Fig. \ref{fig:slope}b. Most of the data points (in particular the points taken for AD \#2 at B = 13.5 T used for the fractional oscillations) lie inside the $\pm 10\%$ interval highlighted by the shaded orange band. This shows that the oscillations are mainly determined by density changes, with the area contribution remaining relatively small. At higher filling factors, such as $\nu = 4$, a larger discrepancy is observed, suggesting that the change in area induced by the top-gate voltage becomes more significant. 

In conclusion, we have demonstrated that a gate-defined bilayer graphene AD serves as a powerful yet simple platform for measuring fractional charge in the quantum Hall regime. Owing to its simple and versatile design, this platform can be readily extended to other material systems, for instance, for charge measurements in anomalous quantum Hall states in rhombohedral graphene or in twisted transition metal dichalcogenides.

\bibliography{biblio}

\section*{METHODS}

\subsection{Sample fabrication}

The devices were fabricated using a standard van der Waals dry-transfer technique. First, graphene and hBN were mechanically exfoliated from bulk crystals on a SiO2/Si substrate. The desired flakes are identified under an optical microscope and atomic force microscope to check for any flake impurity. The stack was assembled using homemade poly(bisphenol A carbonate)/polydimethylsiloxane (PC/PDMS) to pick up all the flakes at a temperature $\sim$ 90 $^\circ$C. The stack was then transferred onto a doped silicon substrate with a 285 nm thick layer of thermally grown SiO$_2$ by melting the PC at a temperature of $\sim$ 180 $^\circ$C. The measured stacks have a 56 nm (45 nm) top hBN layer and a 33 nm (30 nm) bottom hBN layer for AD \#1 (AD \#2).

Device patterning was achieved through multiple steps of electron beam lithography, followed by reactive ion etching and metal deposition. First, the TG contacts were created by depositing an 18 nm Pd layer. For AD \#1 the device geometry was then defined through reactive ion etching, initially using O$_2$, followed by SF$_6$, and then another O$_2$ step. The electrode pattern was subsequently created by etching with SF$_6$ and O$_2$, followed by angled deposition of a 5/100 nm Cr/Au layer with rotation. For AD \#2 instead, the device geometry was then defined through reactive ion etching, initially using O$_2$, followed by CHF$_3$+O$_2$. The electrode pattern was subsequently created by etching with O$_2$, CHF$_3$+O$_2$, followed by angled deposition of a 5/12/60 nm Cr/Pd/Au layer with rotation.

Finally, the TG was refined with 30-second O$_2$ etch steps. During each step, the two-probe resistance between each gate bridge-contact was monitored to ensure that all gates were fully separated while minimizing etching of the hBN and achieving narrow line widths \cite{Ronen2021May}.\\

\subsection{Measurements}

The device was measured in a BlueFors dry dilution refrigerator with a base temperature of $\sim$ 10 mK. Electronic filters are installed on all transport lines to help thermalize electrons. Resistance measurements are conducted using ZI lock-in amplifiers with a 1 V AC voltage excitation at a frequency of 33.333 Hz for AD \#1 and 107.777 Hz for AD \#2 applied to a homemade voltage divider. A Yokogawa GS200 sets the dc voltage bias. The current is measured through a current-to-voltage amplifier (K-tip variable gain transimpedance amplifier), and voltages are amplified at room temperature by 100 using LI-75A low noise voltage preamplifiers, before being sent to the lock-in amplifier. For dc bias measurements, a homemade voltage adder is employed. Yokogawa GS200 is used to apply a dc voltage to each gate. To improve the quality of the ohmic contacts, a positive voltage of $V_\mathrm{Si} = 14.5$ V is applied to the silicon gate for AD \#1 and $V_\mathrm{Si} = 53.2$ V for AD \#2, doping the graphene layer next to the contacts in a highly electron-doped region.

\textbf{\begin{center}Author contributions\end{center}}
M.B., M.D.L., and Z.Z. conceived the project. M.B. supervised the project. M.D.L. fabricated the devices with inputs from Z.Z. and T.F. M.D.L. performed the measurements and analyzed the data with input from E.H. T.L. and S.H.S. performed the theoretical calculations. K.W. and T.T. provided the hBN crystals. M.D.L., E.H., and M.B. wrote the manuscript with input from all authors.

\begin{acknowledgments}
The authors acknowledge B. I. Halperin for detailed discussions and help with the interpretation of the observations. The authors also acknowledge M. Heiblum, and F. Kuemmeth for valuable experimental and theoretical insights. M. B. acknowledges helpful discussions with H-S. Sim., P. Roulleau, and C. R. Dean regarding the interpretation of the data. M.D.L., E.H., and Z.Z. acknowledge funding from SNSF. M.B. acknowledges the support of the SNSF Eccellenza grant No. PCEGP2\_194528, and support from the QuantERA II Programme that has received funding from the European Union’s Horizon 2020 research and innovation program under Grant Agreement No 101017733. K.W. and T.T. acknowledge support from the JSPS KAKENHI (Grant Numbers 20H00354 and 23H02052) and World Premier International Research Center Initiative (WPI), MEXT, Japan. M.D.L. acknowledges Blender.
\end{acknowledgments}

\textbf{\begin{center}Data availability\end{center}}
The data supporting the findings of this study are available from the corresponding author upon reasonable request.

\onecolumngrid
\newpage

\clearpage

\section*{Supplementary Materials}
\subsection{Devices Characterization}
\label{sec:characterization}

\begin{figure*}[htb!]
\renewcommand{\thefigure}{S1}
 \includegraphics[width = \textwidth]{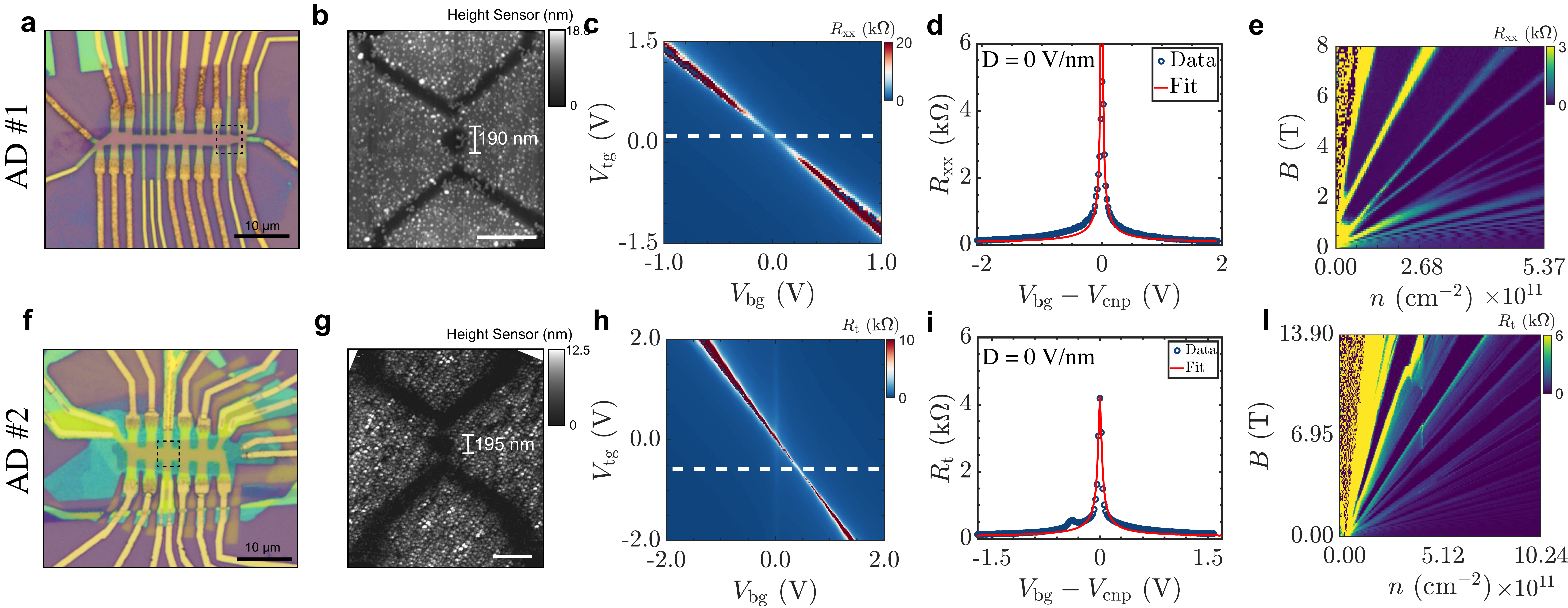}
 \begin{center}
 \caption{\textbf{Devices characterization.} \textbf{a, f}, Optical microscope image of the devices AD \#1 and AD \#2, respectively. \textbf{b, g}, AFM height sensor image of the AD \#1 and AD \#2. The AD diameters are $D_\mathrm{AD\#1} = 190 $ nm and $D_\mathrm{AD\#2} = 195 $ nm. The scale bars are 400 nm. \textbf{c, g}, Longitudinal resistances measured outside the AD at $B = 0$ T as a function of the BG and TG voltage for AD \#1, while for AD \#2 it has been measured across the AD. \textbf{d, i}, Longitudinal resistance at zero displacement field for both devices as a function of the BG in blue and best fit in red. \textbf{e, l}, Longitudinal resistance and transmitted resistance fan diagram
for AD \#1 and AD \#2 as a function of the carrier density $n$ and the magnetic field $B$.} 
 \label{fig:supp_1}
 \end{center}
\end{figure*}

Figure \ref{fig:supp_1}a,e shows optical microscope images of the devices. The black dashed square marks the location of the AD. Figures \ref{fig:supp_1}b,f present AFM height sensor images of the AD \#1 and AD \#2. From the height sensor, we measure the etched AD diameters to be $D_\mathrm{AD\#1} = 190 $ nm and $D_\mathrm{AD\#2} = 195 $ nm. \\

From the measurements Hall conductance dependence, measured outside the AD, we extract the BG capacitance per unit area $C^\mathrm{1}_\mathrm{bg} = 605 \pm 2$ $\mu$Fm$^{-2}$ and from the slope in Fig. \ref{fig:supp_1}c the TG Capacitance $C^\mathrm{1}_\mathrm{tg} = 451 \pm 5$ $\mu$Fm$^{-2}$ for the first device AD \#1.
While for the second device we measure $C^\mathrm{2}_\mathrm{bg} = 805 \pm 2$ $\mu$Fm$^{-2}$ and from the slope in Fig. \ref{fig:supp_1}h the TG Capacitance $C^2_\mathrm{tg} = 549 \pm 5$ $\mu$Fm$^{-2}$.

Following the procedure in \cite{Kim2024Nov} the measured $R_\mathrm{xx}$ can be fitted with the following equation

\begin{equation}
 R_\mathrm{xx} = \frac{L/W}{\mu \sqrt{(C_\mathrm{bg}(V_\mathrm{bg} - V_\mathrm{cnp}))^2 + (n_0 e)^2)}}
\end{equation}

\noindent with $V_\mathrm{cnp}$ the voltage corresponding to the charge neutrality point, $L$ and $W$ the length between two contacts and the width of the device respectively, $\mu$ the mobility at high density, and $n_0$ the charge background fluctuations, which are proportional to the intrinsic doping. From the best fit, shown in Fig. \ref{fig:supp_1}d we find a mobility $\mu^1 \simeq 180,000$ cm$^{-2}$V/s and a charge impurity of $n^1_0 = 1.1 \cdot 10^{10}$ cm$^{-2}$ for the first device and $\mu^2 \simeq 240,000$ cm$^{-2}$V/s and a charge impurity of $n^2_0 = 6 \cdot 10^{8}$ cm$^{-2}$ for the second device. We believe that the absence of fractional quantum Hall states at a field of B = 8 T is due to the relatively high density of charge impurities. \\

The vertical resistive line in Fig. \ref{fig:supp_1}h arises from the side gates, since the resistance is measured across the AD.\\

Figure \ref{fig:supp_1}e,l show the magneto-transport measurements for both devices. In BLG, Landau levels exhibit a fourfold degeneracy due to spin and valley degrees of freedom. Additionally, the two lowest orbital levels share the same energy due to the LL energy dispersion, given by $E = \hbar\omega_\mathrm{c}\sqrt{N(N-1)}$ where $\omega_\mathrm{c}$ is the frequency of the cyclotron orbit. As the magnetic field increases, the Landau level degeneracy is lifted first for the spin, followed by valley splitting, and finally by the orbital number splitting \cite{Hunt2017Oct, Li2018Jan, Kumar2025Aug}.\\

\begin{figure}[htb!]
\renewcommand{\thefigure}{S2}
 \includegraphics[width = 0.5\textwidth]{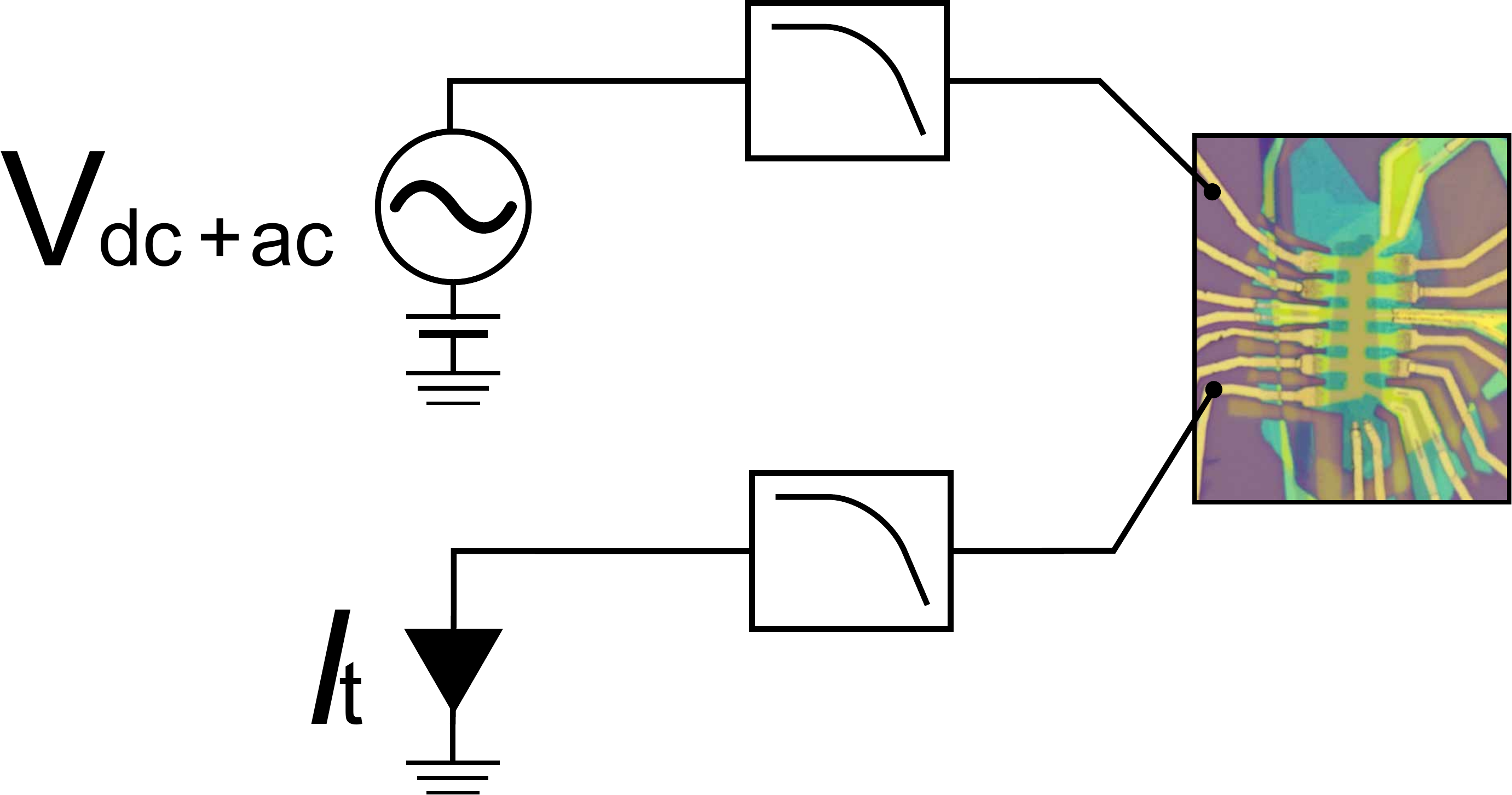}
 \begin{center}
 \caption{\textbf{Current measurement.} Layout of the wiring for the conductance measurements. The measured current accounts for a voltage drop across the sample and also an additional contribution from the voltage drop across the dc lines and the low-pass filters.} 
 \label{fig:supp_1b}
 \end{center}
\end{figure}

In our setup, a voltage bias is applied to the device while the current is measured. The measured current, however, includes not only the voltage drop across the sample but also an additional contribution from the voltage drop across the dc lines and low-pass filters, as illustrated in Fig. \ref{fig:supp_1b}. Because these filtering elements introduce an extra series resistance, the total voltage drop is larger, resulting in a reduced measured current. To account for this effect, we correct the current by applying a multiplicative factor when extracting the conductance. This factor is chosen such that the $\nu = 2$ quantum Hall plateau is properly quantized. In our case, the correction factor is $f = 1.02$.

\newpage

\subsection{Device Tuning}
\label{sec:tuning}

\subsubsection*{Silicon Gate Characterization}

\begin{figure*}[htb!]
\renewcommand{\thefigure}{S3}
 \includegraphics[width = 0.6\textwidth]{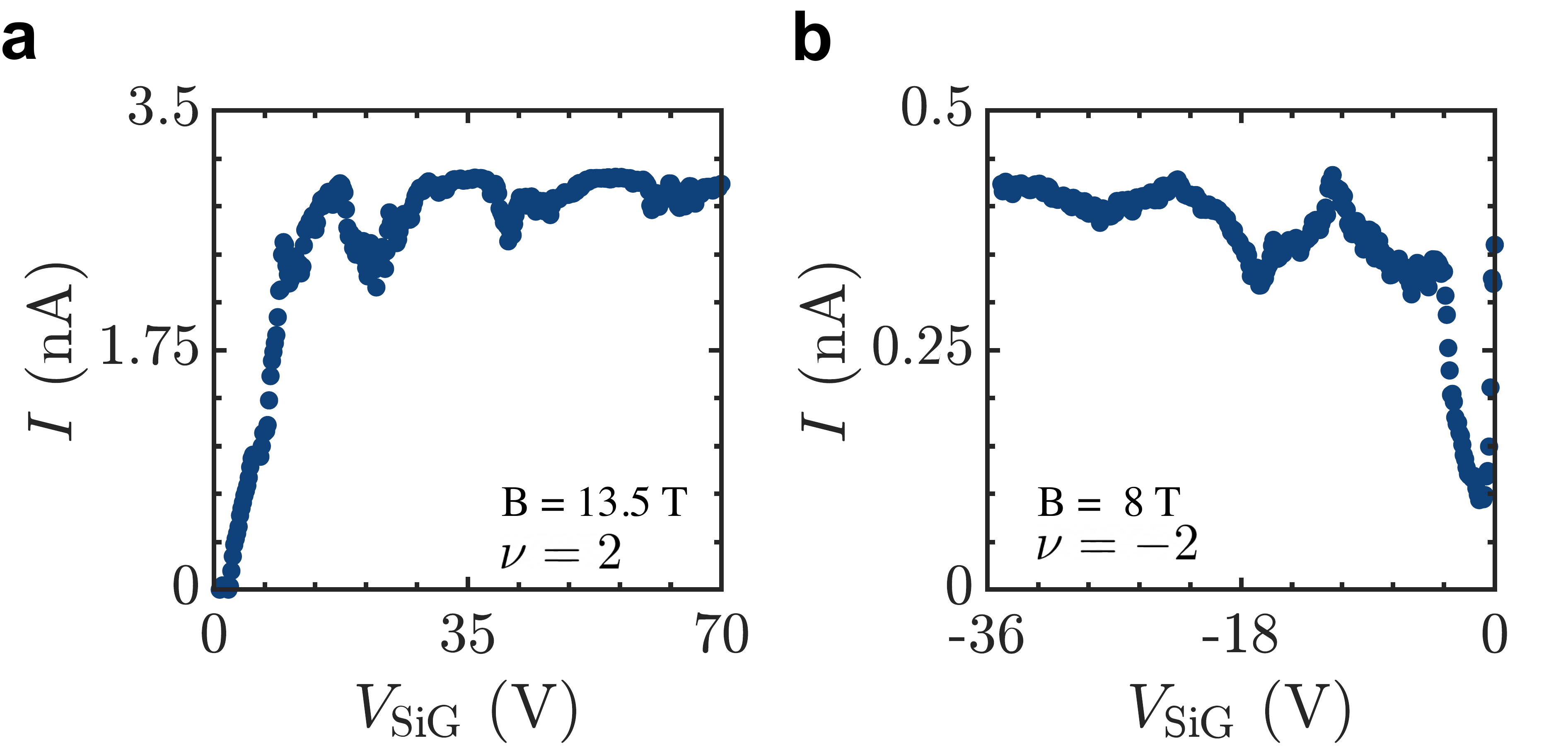}
 \begin{center}
 \caption{\textbf{Silicon gate characterization.} Measured current $I$ as a function of the silicon gate $V_\mathrm{SiG}$ for the electron-side at $B = 13.5$ T (\textbf{a}) and the hole-side at $B = 8$ T (\textbf{b}). For the electron-side at $B = 13.5$ T $V_\mathrm{SiG} = 53.2$ V and for the hole-side at $B = 8 $T $V_\mathrm{SiG} = -46$ V. All the measurements have been performed in AD \#2} 
 \label{fig:supp_2a}
 \end{center}
\end{figure*}

The graphene that extends outside the bottom graphite gate near the ohmic contacts is locally doped using the global silicon back gate. For each magnetic field $B$ the voltage applied to the silicon gate, $V_\mathrm{SiG}$, is chosen by monitoring the transmitted current $I$ on a quantum Hall plateau, where all the current is transmitted. At specific $V_\mathrm{SiG}$ values, the transmitted currents saturate, as shown in Fig. \ref{fig:supp_2a}a-b for the electron-side at $B = 13.5$ T and the hole-side at $B = 8$ T, respectively. We fix the silicon gate $V_\mathrm{SiG}$ at the center of these plateau values. For the electron-side at $B = 13.5$ T $V_\mathrm{SiG} = 53.2$ V and for the hole-side at $B = 8$ T $V_\mathrm{SiG} = -46$ V.

\subsubsection*{Side Gates Characterization}

The side gates enable control over the interaction between the extended edges and the AD–bound states. Applying a negative voltage to the side gates brings the edges closer together, thereby allowing tunneling. 

\begin{figure*}[htb!]
\renewcommand{\thefigure}{S4}
 \includegraphics[width = 0.8\textwidth]{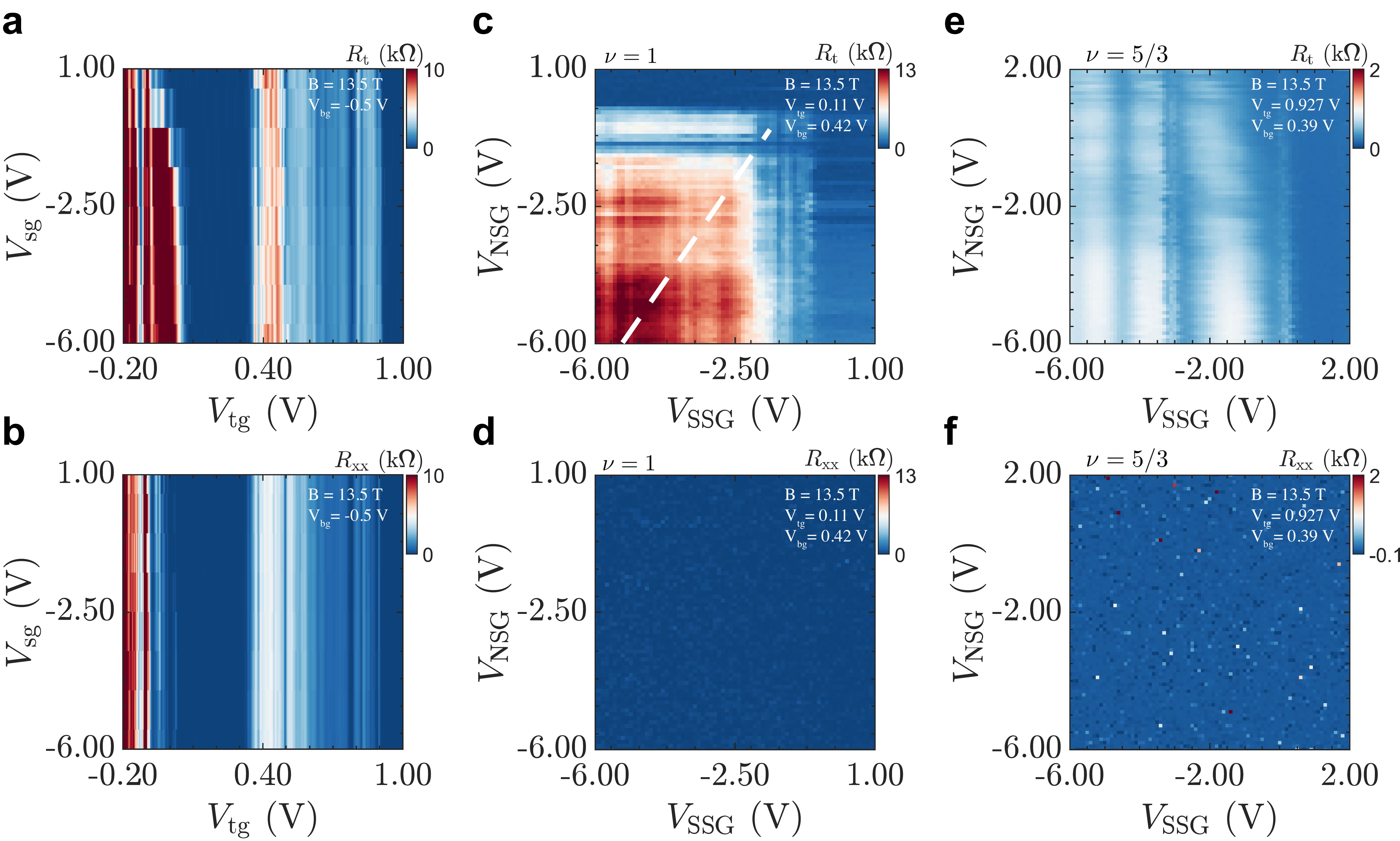}
 \begin{center}
 \caption{\textbf{Side gates characterization}\textbf{a-b}, Transmitted ($R_\mathrm{t}$) and longitudinal (($R_\mathrm{xx}$) resistances in function of the top gate voltage $V_\mathrm{tg}$ and the side gate voltage $V_\mathrm{sg}$ at a BG of $V_\mathrm{bg} = -0.5$ V. For this measurement the side gates were swept symmetrically, $V_\mathrm{SSG} = V_\mathrm{NSG} = V_\mathrm{sg}$. \textbf{c-d}, Transmitted ($R_\mathrm{t}$) and longitudinal (($R_\mathrm{xx}$) resistances in function of the south side gate voltage $V_\mathrm{SSG}$ and the north side gate voltage $V_\mathrm{NSG}$ at $V_\mathrm{bg} = 0.42$ V, $V_\mathrm{tg} = 0.11$ V corresponding to the edge of the $\nu = 1$ plateau. The white dotted line in panel \textbf{c} indicates the voltages at which the side gates are symmetric. \textbf{e-f}, Transmitted ($R_\mathrm{t}$) and longitudinal (($R_\mathrm{xx}$) resistances in function of the south side gate voltage $V_\mathrm{SSG}$ and the north side gate voltage $V_\mathrm{NSG}$ at of $V_\mathrm{bg} = 0.39$ V, $V_\mathrm{tg} = 0.927$ V corresponding to the $\nu = 5/3$ plateau. All the measurements have performed with AD \#2 and at $B = 13.5$ T.} 
 \label{fig:supp_2b}
 \end{center}
\end{figure*}

A first characterization consists of fixing the BG voltage and measuring the transmitted resistance ($R_\mathrm{t}$) and longitudinal resistance ($R_\mathrm{xx}$) as functions of the top gate voltage $V_\mathrm{tg}$ and side gate voltage $V_\mathrm{sg}$, as shown in Fig. \ref{fig:supp_2b}a–b. For this measurement, the north side gate (NSG) and the south side gate (SSG) were swept symmetrically, $V_\mathrm{SSG} = V_\mathrm{NSG} = V_\mathrm{sg}$. The $\nu = 1$ plateau narrows as the side gate voltage decreases (Fig. \ref{fig:supp_2b}a), while the plateau measured outside the AD remains constant (Fig. \ref{fig:supp_2b}b). This confirms that the side gates are functioning properly.

However, due to fabrication imperfections, the side gates may not be perfectly symmetric with respect to the AD. At each filling factor, the side gates are tuned to identify the symmetric condition at which the downstream and upstream edges are equally coupled to the AD-bound states. To achieve this, the bulk filling factor is fixed at the edge of the plateau while the NSG and SSG are swept simultaneously. Examples are shown in Fig. \ref{fig:supp_2b}c–d for $\nu = 1$ and Fig. \ref{fig:supp_2b}e–f for $\nu = 5/3$.

For $\nu = 1$, we observe that both gates pinch almost symmetrically, similar to the behavior in quantum dot devices. In this case, the white dotted line indicates the voltages at which the side gates are symmetric. However, for $\nu = 5/3$, one gate exhibits a much stronger effect than the other. We attribute this asymmetry to localized impurities near one gate, which can locally trap electrons. In such cases, we rely on the symmetric diagonal identified for $\nu = 1$. In both cases, the longitudinal resistance $R_\mathrm{xx}$ remains constant.

\begin{figure*}[htb!]
\renewcommand{\thefigure}{S5}
 \includegraphics[width = 0.6\textwidth]{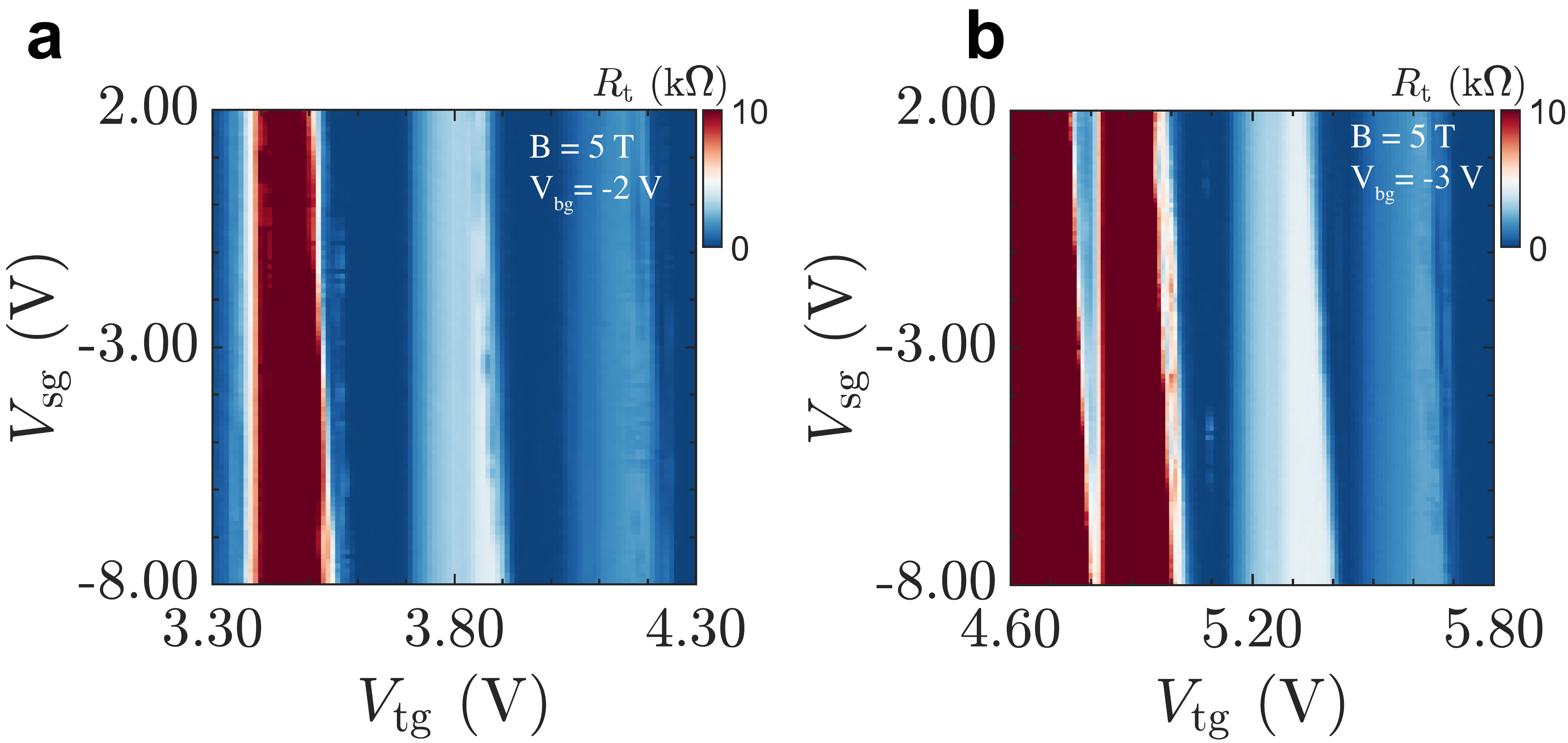}
 \begin{center}
 \caption{\textbf{Side gates at different BG values} \textbf{a-b}, Transmitted resistance $R_\mathrm{t}$ in function of the top gate voltage $V_\mathrm{tg}$ and the side gate voltage $V_\mathrm{sg}$ at a BG of $V_\mathrm{bg} = -2 $V (\textbf{a}) and $V_\mathrm{bg} = -3 $V (\textbf{b}). The TG covers a range from $\nu = 1$ to $\nu = 4$. A more negative bottom gate voltage helps pinch the edges. The side gates were swept symmetrically, $V_\mathrm{SSG} = V_\mathrm{NSG} = V_\mathrm{sg}$. All the measurements have been performed in AD \#2.} 
 \label{fig:supp_2c}
 \end{center}
\end{figure*}

Finally, the voltage applied to the BG also controls the pinching of the side gates, as shown in \ref{fig:supp_2c}, with a more negative bottom gate voltage that helps pinching the edges.

\newpage

\subsection{Discussion on the error estimation of the oscillation periods}
\label{sec:error}

For each oscillation, the two-dimensional Fast Fourier Transform provides information about the periodicity of the oscillations. However, the 2D-FFT can present artifacts due to the fact that the oscillations are not infinite but are abruptly truncated at the limits of the magnetic field range and gate voltage range.
To reduce these artifacts, a windowing function is applied to the curves before performing the 2D-FFT:

$$
\tilde{S}(f_B, f_V) = \mathcal{F}_{2D}\left[ W(B,V) \cdot S(B,V) \right]
$$

\noindent where $S(B,V)$ is the signal, and $W(B,V)$ is the windowing function, in our case the Hamming function.\\

The characteristic period is usually extracted from the highest peak of the 2D-FFT. In some cases, such as for some oscillations in the FQHE, this peak is not objectively distinguishable.
In these cases, the periodicity is estimated from the spacing between multiple consecutive peaks in the original conductance signal. These periods obtained from the pijama plot are then compared with the peaks present in the 2D-FFT spectrum for consistency, and used to identify the correct frequency. \\

Finally, a crucial part of the analysis is the choice of the error on the extracted frequency. The discrete steps in gate voltage ($dV$) and magnetic field ($dB$) are taken as the minimum units of error, since they correspond to the minimal observable resolution in the FFT:

$$
\delta f_V \sim \frac{1}{N_V \, dV}, 
\qquad 
\delta f_B \sim \frac{1}{N_B \, dB}
$$

\noindent where $N_V$ and $N_B$ are the number of sampled points in gate and field, respectively. In addition, an error term is added in quadrature to account for the uncertainty in determining the correct frequency when the peak is broad or not clearly defined. Most of the measurements used in the main text have been carried out with a field discrete step of $dB = 1$ mT, lower limit of our magnet power supply, and a gate voltage step of $dV = 0.1$ mV; the scans contain around $N_B = N_V = 101$ points. \\

The largest source of uncertainty arises from the finite size of the FQHE plateaus, which typically extend over a limited range ($\sim 200$ mT and $\sim 50$ mV). This means that only a limited number of oscillations can be observed in both the magnetic field and gate voltage before the system leaves the fractional plateau. In practice, no more than about 10 oscillations can be resolved; after this point, the bulk no longer remains in the fractional quantum Hall plateau, making it difficult to extract periodicity with high precision. This limitation directly increases the error on the extracted frequency, since the frequency resolution is inversely proportional to the available range. Thus, the shorter plateau length in the FQHE compared to the IQHE is the dominant factor that limits the precision of the periodicity analysis. \\

\newpage

\subsection*{Oscillations in the IQH}
\label{sec:IQH}

\subsubsection*{AD \#1}

\begin{figure*}[htb!]
\renewcommand{\thefigure}{S6}
 \includegraphics[width = 0.8\textwidth]{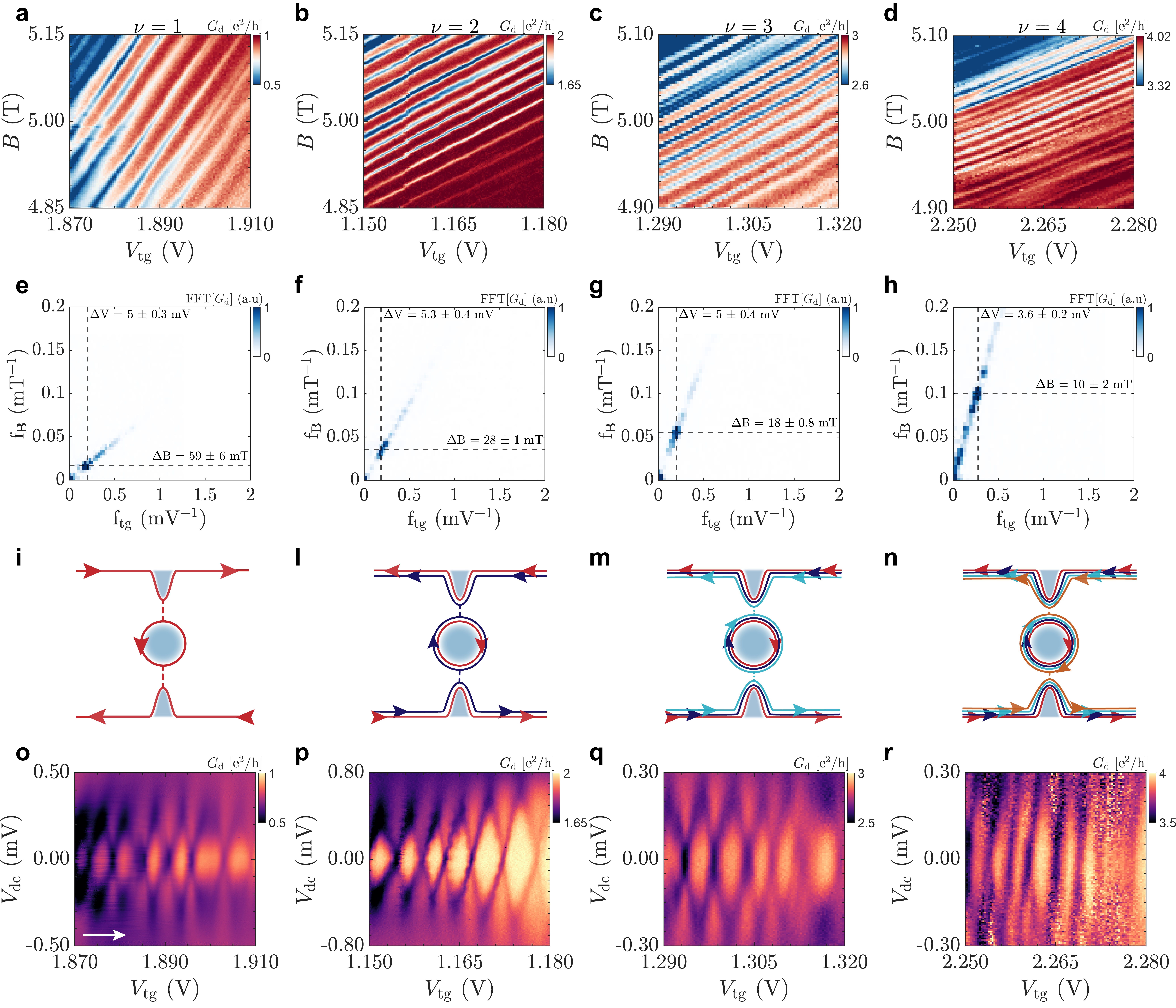}
 \begin{center}
 \caption{\textbf{IQHE oscillations for AD \#1.} \textbf{a-d}, Diagonal conductance $G_\mathrm{d}$ oscillations as a function of the TG voltage and magnetic field for filling factors $\nu = 1-4$. \textbf{e-h}, 2D-FFT of the oscillations. \textbf{g-n} Schematic of the tunneling edge. \textbf{o-r}, Diagonal conductance $G_\mathrm{d}$ oscillations as a function of TG voltage and dc bias, displaying the characteristic Coulomb diamonds. The white arrow in panel \textbf{o} indicates the AD-bound electrons increasing direction.} 
 \label{fig:supp_3a}
 \end{center}
\end{figure*}

Figure \ref{fig:supp_3a} shows the diagonal conductance $G_\mathrm{d}$ oscillations as a function of the TG voltage and magnetic field for filling factors $\nu = 1-4$ (a-d) and their 2D-FFT (e-h) for the first device AD \#1.\\

All oscillations display Coulomb diamond behavior as shown in panels Fig. \ref{fig:supp_3a}o-r. The white arrow in panel (o) indicates the direction of increasing number of electrons bound to the AD. For $\nu = 2$, we observe that the excitation energy increases as the TG voltage increases. This is likely due to a change in the AD potential slope, which results in a higher charge carrier velocity and thus a larger energy gap. For each filling factor, we extract a charging energy, which is on the order of $300-600$ $\mu$eV and depends on the energy level spacing. It is highest at $\nu = 2$, followed by $\nu = 4$, $\nu = 1$, and finally $\nu = 3$. This trend aligns with the Landau level structure in bilayer graphene, where the Landau level gaps at $\nu = 4$, and $\nu = 2$, are the largest \cite{Martin2010Dec}. To estimate the energy level spacing without considering electron-electron interactions, we can use the expression $\delta \epsilon = 2\hbar v/D_{AD}$ with $\hbar$ is the reduced Planck's constant and $v = -\frac{1}{eB} \frac{\partial V}{\partial r}$ is the drift velocity of the charge carriers on the AD potential $V$. In graphene interferometers, a typical value for the drift velocity is $v = 10^5 $ m/s \cite{Ronen2021May, Deprez2021May, Werkmeister2025Apr, Samuelson2024Mar, Kim2024Nov}. Assuming an AD diameter $D_\mathrm{AD}$ of 300 nm, the energy gap is $\delta \epsilon \simeq 450$ $\mu$eV. This estimated energy gap aligns well with experimental observations without the need for electron-electron interactions. 

\newpage

\subsubsection*{AD \#2}

\begin{figure*}[htb!]
\renewcommand{\thefigure}{S7}
 \includegraphics[width = 0.9\textwidth]{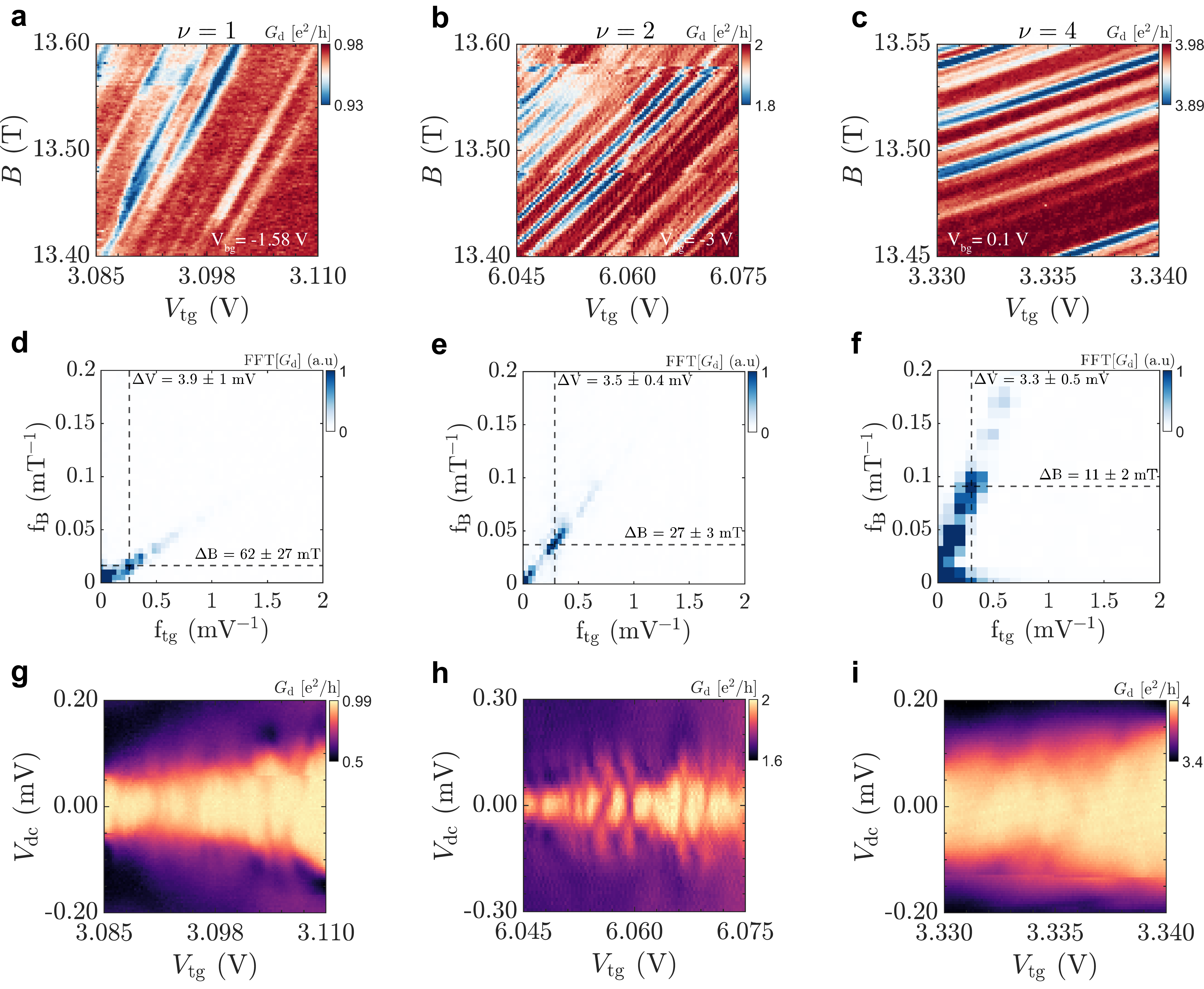}
 \begin{center}
 \caption{\textbf{IQHE oscillations for AD \#2.} \textbf{a-c}, Diagonal conductance $G_\mathrm{d}$ oscillations as a function of the BG voltage and magnetic field for filling factors $\nu = 1,2$ and 4. \textbf{d-f}, 2D-FFT of the oscillations. \textbf{g-i}, Diagonal conductance $G_\mathrm{d}$ oscillations as a function of TG voltage and dc bias.} 
 \label{fig:supp_3b}
 \end{center}
\end{figure*}

\begin{figure*}[htb!]
\renewcommand{\thefigure}{S8}
 \includegraphics[width = \textwidth]{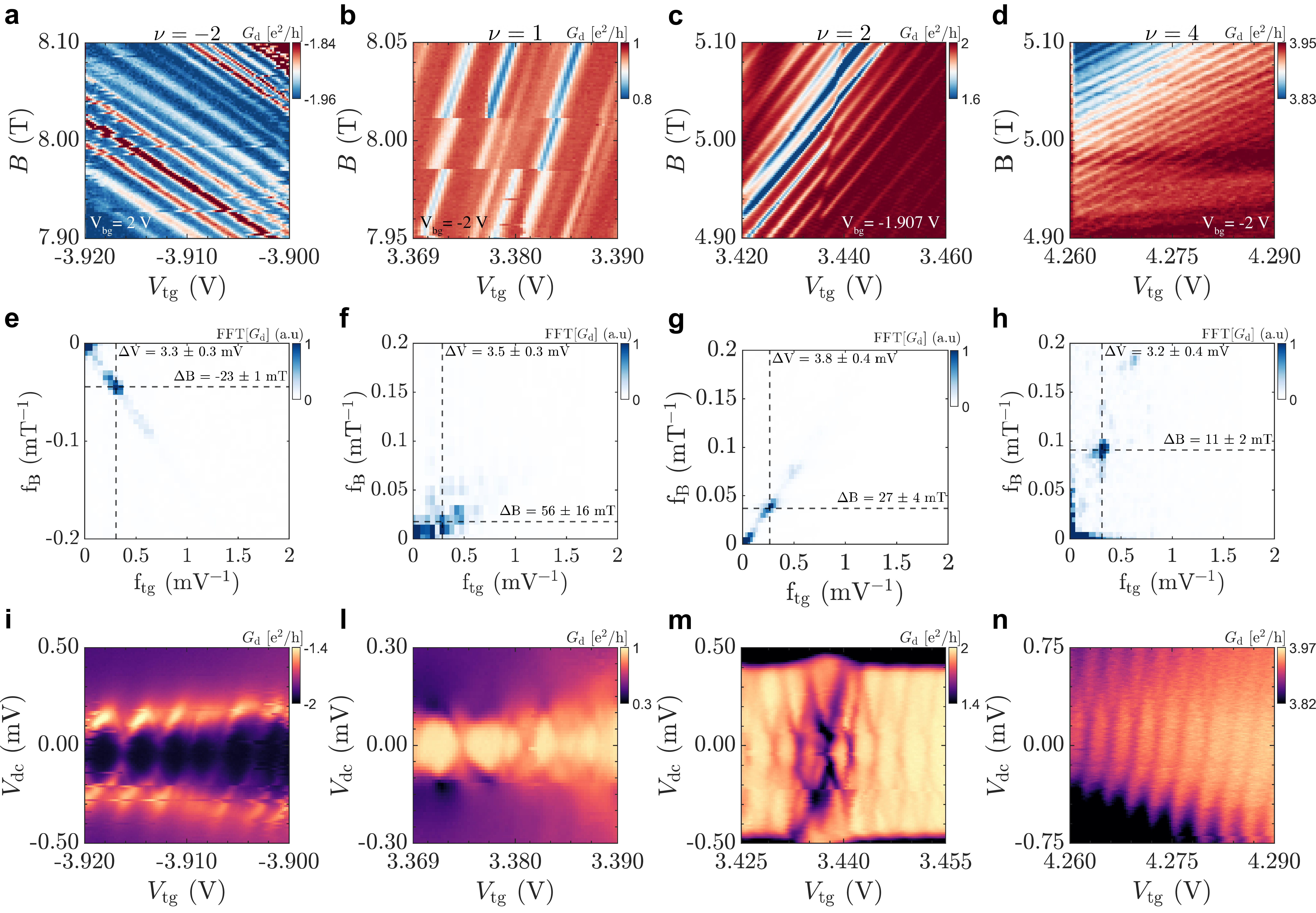}
 \begin{center}
 \caption{\textbf{Lower field IQHE oscillations for AD \#2.} \textbf{a-d}, Diagonal conductance $G_\mathrm{d}$ oscillations as a function of the TG voltage and magnetic field for filling factors $\nu = -2,1, 2$ and 4. \textbf{e-h}, 2D-FFT of the oscillations, the insert schematically shows the interfering edge. \textbf{i-n}, Diagonal conductance $G_\mathrm{d}$ oscillations as a function of TG voltage and dc bias. The measurements for $\nu = 4$ at $B = 5$ T have been taken at a temperature of T = 300 mK.} 
 \label{fig:supp_3c}
 \end{center}
\end{figure*}

Figure \ref{fig:supp_3b} shows the diagonal conductance $G_\mathrm{d}$ oscillations as a function of the TG voltage and magnetic field for filling factors $\nu = 1, 2$ and 4 (a–c) and their corresponding 2D-FFT spectra (d–f), measured for the second device (AD \#2) at $B = 13.5$ T. From the Source-Drain measurements (g-i) we can extract the charging energy. It can be observed that the charging energies are much smaller at $B = 13.5$ T compared to lower magnetic fields (Fig. \ref{fig:supp_3c}), as expected from the scaling $\Delta \epsilon \sim 1/B$. Furthermore, for $\nu = 1$ and $\nu = 4$, it appears that the device may not be operating in the single-level hopping regime ($k_b T \ll \Delta \epsilon$), as Coulomb diamonds are difficult to resolve. In these cases, transport is likely dominated by multilevel hopping ($k_b T \geq \Delta \epsilon$). This can also explain the observed quasiperiodic oscillations.\\

\newpage 

Figure \ref{fig:supp_3c} shows the diagonal conductance $G_\mathrm{d}$ oscillations as a function of the TG voltage and magnetic field for filling factors $\nu = -2, 1, 2,$ and $4$ (a–d), together with their corresponding 2D-FFT spectra (e–h), measured for the second device (AD \#2). On the hole side, at $\nu = -2$, the oscillations display an opposite slope, as expected from density oscillations. In this case, we changed the contact configuration to account for the opposite chirality of the edge states.

\newpage

\subsection{Oscillation in Bottom Gate}
\label{sec:BG}

As mentioned in the main text, another way to achieve oscillations is by varying the BG while keeping the TG constant. However, the drawback of this method is that the AD potential and the side gates trenches are primarily controlled by the BG. Consequently, a small variation in the BG has a much larger effect on the potential landscape compared to a similar variation in the TG. \\

\begin{figure*}[htb!]
\renewcommand{\thefigure}{S9}
 \includegraphics[width = \textwidth]{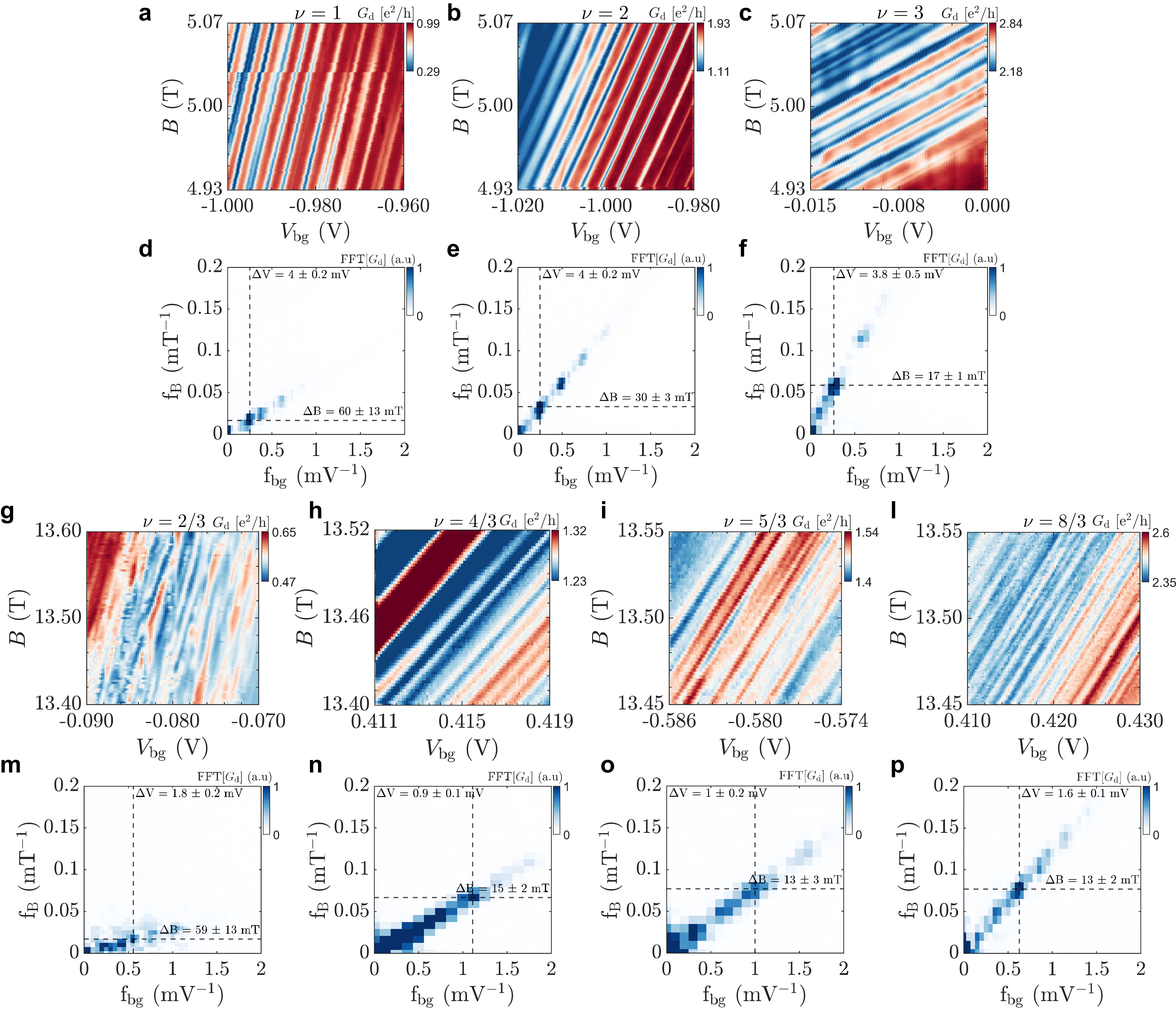}
 \begin{center}
 \caption{\textbf{Oscillations in bottom gate.} \textbf{a-c}, Diagonal conductance $G_\mathrm{d}$ oscillations as a function of the bottom gate voltage and magnetic field for filling factors $\nu = 1-3$ for the first device (AD \#1). \textbf{d-f}, 2D-FFT of the oscillations. \textbf{g-l}, Diagonal conductance $G_\mathrm{d}$ oscillations as a function of the bottom gate voltage and magnetic field for filling factors $\nu = 2/3, 4/3, 5/3$ and 8/3 for the second device (AD \#2). \textbf{d-f}, 2D-FFT of the oscillations.} 
 \label{fig:supp_4a}
 \end{center}
\end{figure*}

Figure \ref{fig:supp_4a}a-c shows oscillations obtained in the BG at $\nu = 1-3$ for the first device (AD \#1), with similar results to those observed in the TG. As shown in Fig. \ref{fig:supp_4a}d-f, the magnetic field period closely matches the one obtained for TG oscillations, as reported in the main text and in Fig. \ref{fig:supp_3a}. However, the BG period differs due to the thinner bottom hBN layer, which results in a larger capacitance. Figure \ref{fig:supp_4a}g-l shows the same measurements performed on AD \#1 for fractional filling factors, $\nu = 2/3, 4/3, 5/3$ and 8/3, with their respective 2D-FFT, Fig. \ref{fig:supp_4a}g-l.\\

\begin{figure*}[htb!]
\renewcommand{\thefigure}{S10}
 \includegraphics[width = 0.9\textwidth]{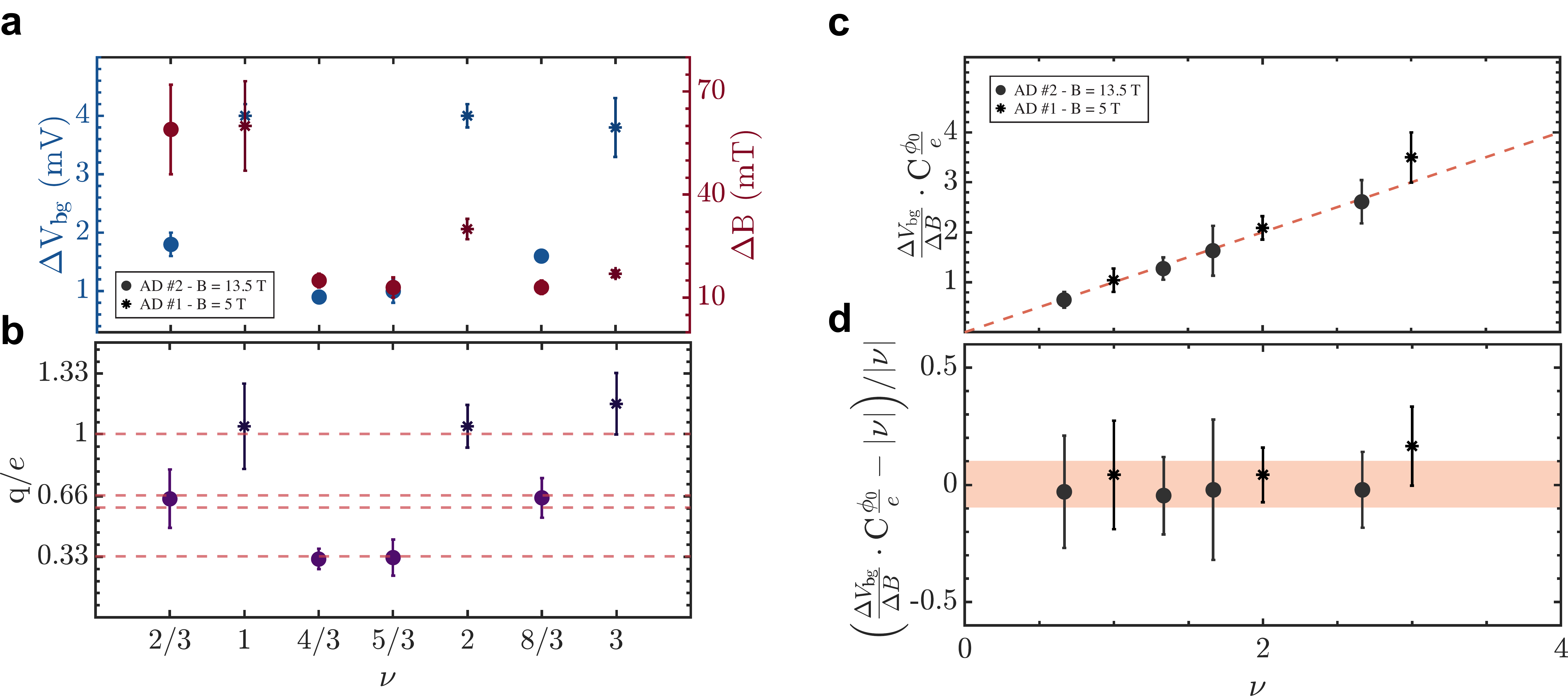}
 \begin{center}
 \caption{\textbf{Charge measurement using the bottom gate.} \textbf{a} Bottom-gate period $\Delta V_\mathrm{bg}$ (blue) and magnetic-field period $\Delta B$ (red) as a function of filling factor $\nu$ for oscillations measured at different magnetic fields and on two devices. \textbf{b}, Quasiparticle charge as a function of filling factor, computed using Eq. \ref{eq:charge}.
\textbf{c}, Ratio of the bottom-gate period to the magnetic-field period, normalized by the bottom-gate capacitance per unit area and $\phi_0/e$, as a function of the filling factor $\nu$. The dotted orange lines are guides to the eye. The oscillations exhibit a slope corresponding to the filling factor $\nu$. \textbf{d}, Relative error of the ratio of the bottom-gate period to the magnetic-field period minus the filling factor. The shaded orange band marks the $\pm 10\%$ interval.} 
 \label{fig:supp_4b}
 \end{center}
\end{figure*}

Figure \ref{fig:supp_4b} reports the periods and the charge quasiparticle charge as a function of filling factor. As for the observations for TG oscillations, we find that the extracted charge follows the same values as the TG. Figure \ref{fig:supp_4b}c-d shows the ratio of the bottom-gate period to the magnetic-field period, normalized by the bottom-gate capacitance per unit area and $\phi_0/e$, as a function of the filling factor $\nu$ with its relative error, respectively. The dotted orange lines are guides to the eye. As for the TG the oscillations exhibit a slope corresponding to the filling factor $\nu$. 

\newpage

\subsection{Oscillation in Area}
\label{sec:Area}

Another way to induce gate oscillations is by modifying the AD potential, which in turn changes the AD area. This can be achieved by sweeping both the TG and the BG simultaneously while maintaining a constant bulk filling factor. In fact, the AD potential is mainly controlled by the BG, if the BG voltage decreases (TG voltage increases to keep a constant bulk filling factor), the AD potential will increase. 

\begin{figure*}[htb!]
\renewcommand{\thefigure}{S11}
 \includegraphics[width = 0.9\textwidth]{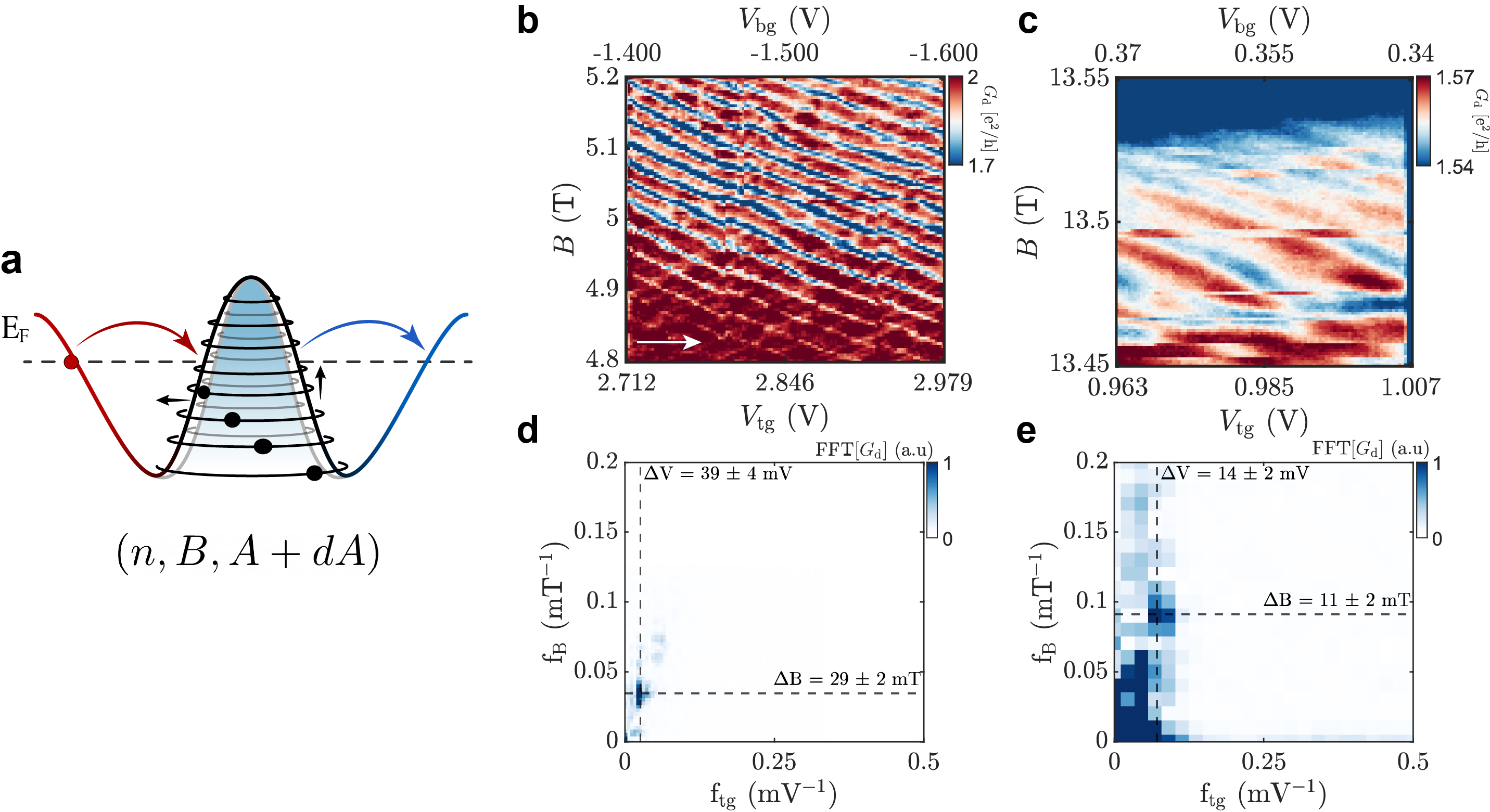}
 \begin{center}
 \caption{\textbf{Oscillations in the AD area.} \textbf{a} Schematic of the tunneling when changing the AD area. Diagonal conductance $G_\mathrm{d}$ oscillations as a function of the area and magnetic field for filling factor $\nu = 2$ in AD \#1 (\textbf{b}) and for $\nu = 5/3$ in AD \#2 (\textbf{c}). \textbf{d-e}, 2D-FFT of the oscillations. The white arrow in panel \textbf{b} indicates the area increasing direction.} 
 \label{fig:supp_5}
 \end{center}
\end{figure*}

As the AD potential increases, so does its area, and to preserve a constant flux, the bound edge states shift to higher energy. An oscillation occurs when the chemical potential of an AD-bound state aligns with the source chemical potential, as schematically illustrated in \ref{fig:supp_5}a.\\

The equation for constant density lines is given by
\begin{equation}
 V_\mathrm{tg} = \alpha V_\mathrm{bg} + \beta
\end{equation}
\noindent where $\beta = \frac{e^2\nu B}{hC_\mathrm{tg}} $ defines the filling factor. The value $\alpha$ is the ratio of the BG and TG capacitance; for AD \#1 we have used $\alpha = -1.34$ and for AD \#2 $\alpha = -1.465$.\\

Figure \ref{fig:supp_5}b-c illustrates the area oscillations for $\nu = 2$ in AD \#1 (b) and for $\nu = 5/3$ in AD \#2. The oscillations exhibit an opposite slope compared to density oscillations due to the opposite direction in which the energy levels move with respect to the increasing direction of the area, as indicated by the white arrow in Fig. \ref{fig:supp_5}b

The 2D-FFT analysis in Fig. \ref{fig:supp_5}d-e reveals a magnetic field period comparable to that observed in density oscillations, while the gate voltage period appears larger with a periodicity of $\Delta \mathrm{V} \simeq 39 $ mV, which is almost nine times larger than the one obtained when changing the density. This suggests that the energy levels of the AD-bound states shift up by at least one level every nine electron jumps, which is consistent with a 5\% overestimation of the electron charge using only the TG voltage. To confirm this theory, we performed oscillations using only BG (section \ref{sec:BG}) that show an estimate of the tunneling charge of about 5\% less than the electron charge. 

\newpage

\subsection{Outer Edge at $\nu = 2$}
\label{sec:outer_edge}

As the constriction filling factor continues to decrease, the inner edge can become completely reflected, allowing the outer edge to start interfering. In this case, we expect the magnetic field oscillation period to be $\phi_0/\nu_\mathrm{int}$ with $\nu_\mathrm{int} = \nu_\mathrm{b} -1$. \\

\begin{figure*}[htb!]
\renewcommand{\thefigure}{S12}
 \includegraphics[width = \textwidth]{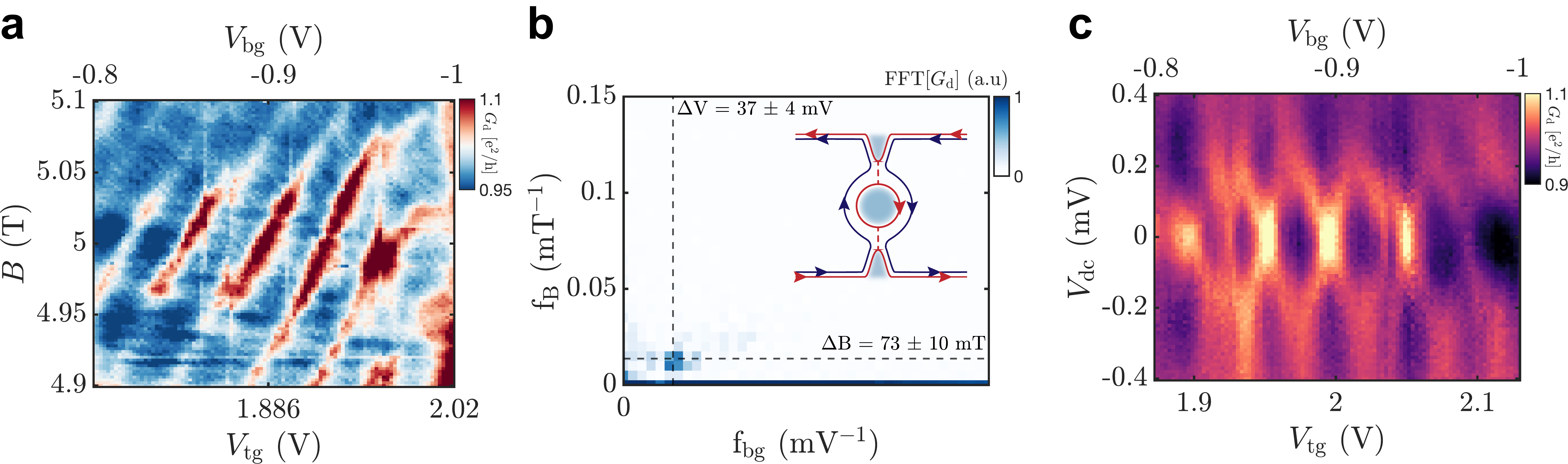}
 \begin{center}
 \caption{\textbf{Oscillations in the outer edge of $\nu = 2$ } \textbf{a} Diagonal conductance $G_\mathrm{d}$ oscillations as a function of the AD area magnetic field for the outer edge of filling factor $\nu = 2$. \textbf{b}, 2D-FFT of the oscillations. The insert schematically shows the outer edge interfering while the inner one is fully reflected. \textbf{c}, Diagonal conductance $G_\mathrm{d}$ oscillations as a function of the AD area dc bias.} 
 \label{fig:sup_7}
 \end{center}
\end{figure*}

Figure \ref{fig:sup_7}a shows oscillations for the outer edge at a filling factor of 2 ($\nu_\mathrm{b} = 2$, $\nu_\mathrm{int} = 1$). From the 2D-FFT, we extract a magnetic field period of $\Delta \mathrm{B} = 73 \pm 10 $ mT, which is similar to that of the inner edge at filling factor 1. This corresponds to a flux period of $\phi_0$ and an AD diameter of $D = 269 \pm 18 $ nm. Since the oscillations are gate-induced, we cannot directly extract the interfering charge.\\

Source-drain measurements, shown in Fig. \ref{fig:sup_7}c, still display a Coulomb-dominated pattern with a charge excitation energy of $E = 210 \pm 10 $ $\mu$eV.

\newpage

\subsection{Oscillations in the FQHE}
\label{sec:FQHE}

In this section, we report oscillations measured at different fractional filling factors. Each oscillation is identified by the bottom-gate voltage $V_\mathrm{bg}$ at which it was acquired, together with the corresponding displacement field, $D$. Almost all oscillations occur at low displacement field where the orbital number remains the same. For $\nu = 5/3$, oscillations were also taken at larger displacement fields without any detectable variation, suggesting that the displacement field primarily modifies the potential slope of the AD without altering the nature of the oscillations.\\

\begin{figure*}[htb!]
\renewcommand{\thefigure}{S15}
 \includegraphics[width = 0.95\textwidth]{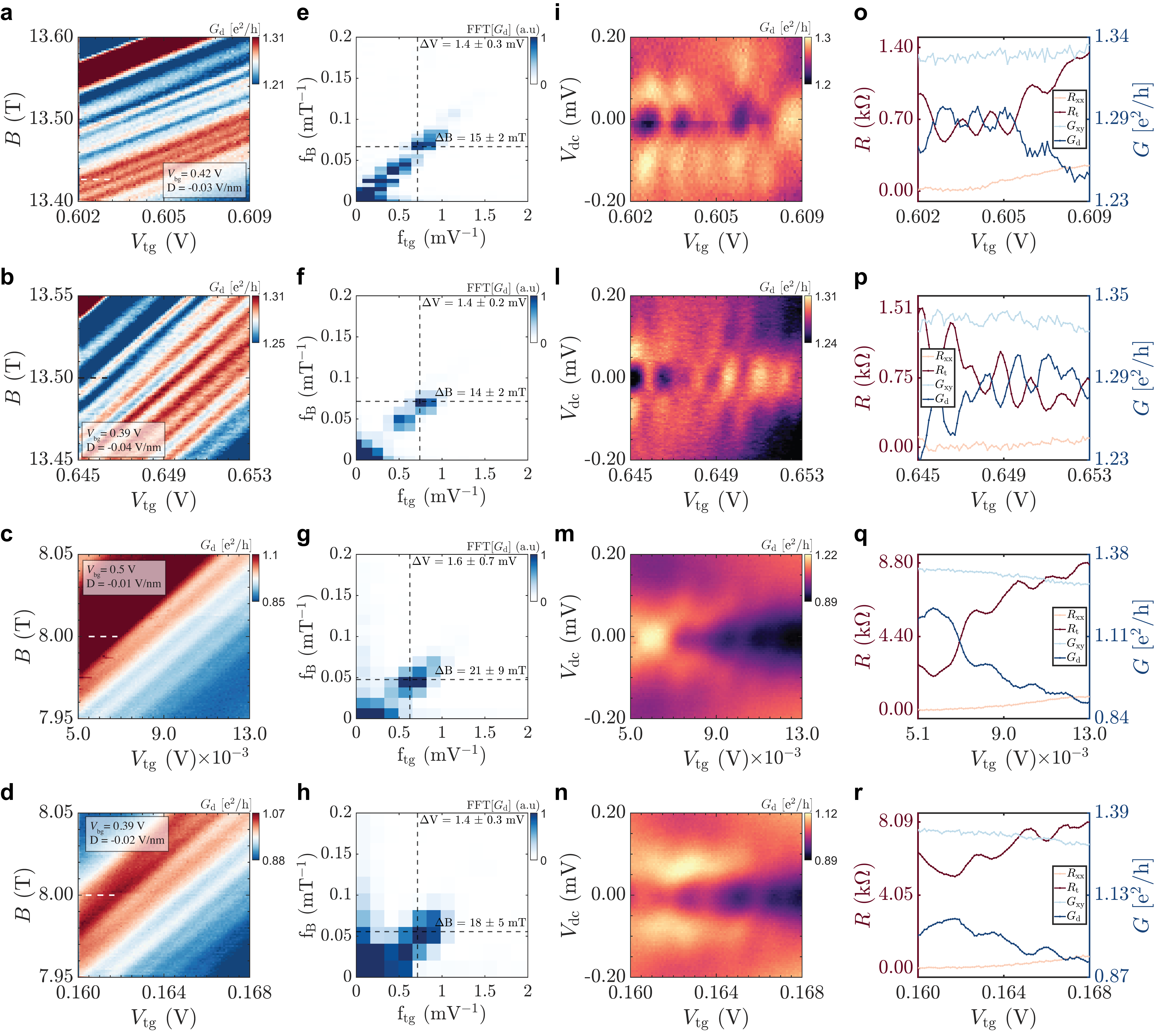}
 \begin{center}
 \caption{\textbf{Oscillations at $\nu = 4/3$.} \textbf{a-d}, Diagonal conductance $G_\mathrm{d}$ oscillations as a function of the TG voltage and magnetic field for filling factor $\nu = 4/3$ at different displacement fields $D$ and at different magnetic fields. \textbf{e-h}, 2D-FFT of the oscillations. \textbf{i-n}, Diagonal conductance $G_\mathrm{d}$ oscillations as a function of TG voltage and dc bias. \textbf{o-r}, Transmitted resistance measured across the AD ($R_\mathrm{t}$ in dark red), longitudinal resistance measured outside the AD ($R_\mathrm{xx}$ in light red), diagonal conductance ($G_\mathrm{d}$ in dark blue) and Hall conductance ($G_\mathrm{xy}$ in light blue) as a function of TG voltage. The oscillations are taken along the dotted lines in their respective panels \textbf{a-d}.} 
 \label{fig:supp_6c}
 \end{center}
\end{figure*}

For each set of oscillations we have also reported the source–drain bias dependence. The latter provides information on the energy gap of the underlying states. From each fractional filling factor, we extract a charging energy on the order of $50 \sim \mu$eV, which is nearly an order of magnitude smaller than that obtained for integer oscillations. This behavior is expected, since fractional states generally exhibit smaller energy gaps. Moreover, the charging energy (or level spacing) scales inversely with both the magnetic field and the quasiparticle charge, $\delta \epsilon \sim -\frac{1}{e^*B}$. As a result, at higher magnetic fields, the energy spacing becomes smaller. At the same time, the reduced quasiparticle charge $e^*$ in fractional states increases the spacing, making fractional oscillations comparatively easier to observe. However, as shown in Fabry–Pérot interferometer and STM experiments \cite{Deprez2021May, Moreau2022Mar}, the bias dependence is not a unique fingerprint of the oscillation mechanism. In some cases, conventional Coulomb diamonds are observed, consistent with the IQHE regime where the AD is filled and tunneling is suppressed. In other cases, inverted diamonds appear, characterized by a conductance minimum at zero bias. Figure \ref{fig:supp_6c}l shows the source–drain bias dependence of the oscillation at $\nu = 4/3$, also discussed in the main text. It can be observed that the zero-bias conductance oscillation exhibits a transition from a peak to a dip. This behavior suggests a phase shift in the oscillation pattern, which is consistent with an Aharonov–Bohm–like mechanism rather than a purely Coulomb-dominated regime. However, due to the limited extent of the fractional plateaus, systematic exploration across the plateau was not possible, preventing a definitive conclusion. \\

Finally, for each oscillation, we also present a line cut as a function of the top-gate voltage. The transmitted resistance across the AD ($R_\mathrm{t}$, dark red), the longitudinal resistance outside the AD ($R_\mathrm{xx}$, light red), the diagonal conductance ($G_\mathrm{d}$, dark blue), and the Hall conductance ($G_\mathrm{xy} = R_0/(R_\mathrm{d} - R_\mathrm{t})$, light blue, with $R_0$ the quantum of resistance) are simultaneously displayed. Oscillations are consistently measured at the center of the fractional plateau, where $R_\mathrm{xx} = 0$ and $G_\mathrm{xy}$ is constant. In this regime, the oscillations in $R_\mathrm{xx}$ and $G_\mathrm{d}$ are fully correlated, confirming that they originate from the AD.\\

\begin{figure*}[htb!]
\renewcommand{\thefigure}{S13}
 \includegraphics[width = 0.95\textwidth]{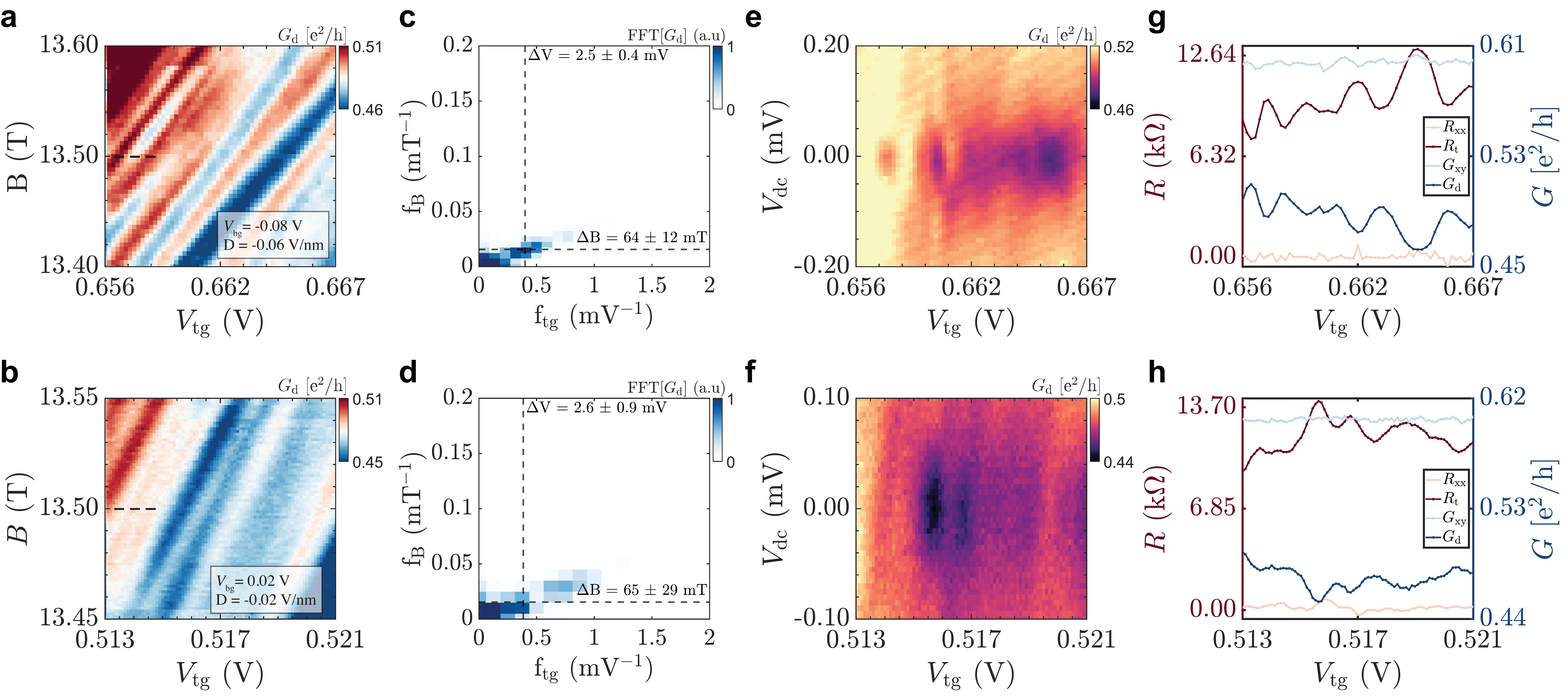}
 \begin{center}
 \caption{\textbf{Oscillations at $\nu = 3/5$.} \textbf{a-b}, Diagonal conductance $G_\mathrm{d}$ oscillations as a function of the TG voltage and magnetic field for filling factor $\nu = 3/5$ at different displacement fields $D$. \textbf{c-d}, 2D-FFT of the oscillations. \textbf{e-f}, Diagonal conductance $G_\mathrm{d}$ oscillations as a function of TG voltage and dc bias. \textbf{g-h}, Transmitted resistance measured across the AD ($R_\mathrm{t}$ in dark red), longitudinal resistance measured outside the AD ($R_\mathrm{xx}$ in light red), diagonal conductance ($G_\mathrm{d}$ in dark blue) and Hall conductance ($G_\mathrm{xy}$ in light blue) as a function of TG voltage. The oscillations are taken along the dotted lines in their respective panels \textbf{a-b}.} 
 \label{fig:supp_6a}
 \end{center}
\end{figure*}

\begin{figure*}[htb!]
\renewcommand{\thefigure}{S14}
 \includegraphics[width = 0.95\textwidth]{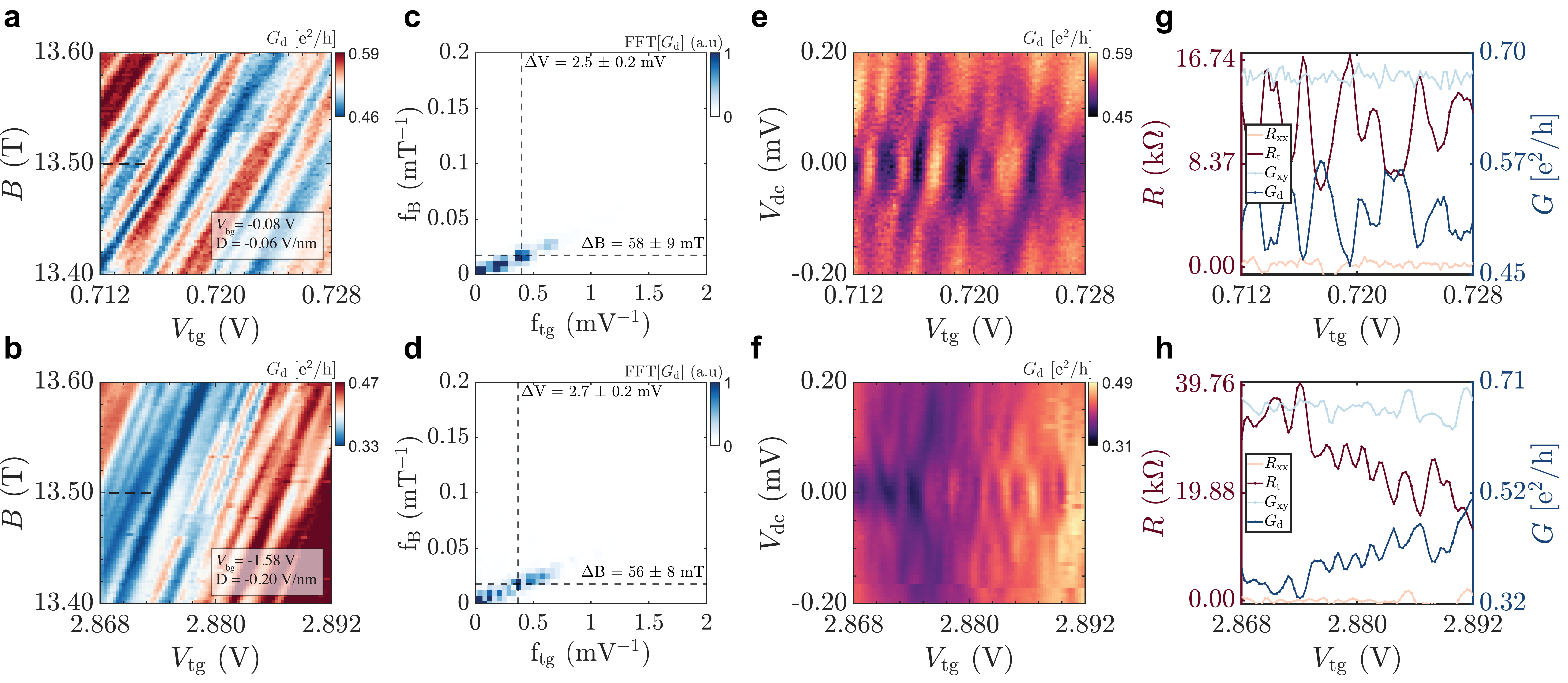}
 \begin{center}
 \caption{\textbf{Oscillations at $\nu = 2/3$.} \textbf{a-b}, Diagonal conductance $G_\mathrm{d}$ oscillations as a function of the TG voltage and magnetic field for filling factor $\nu = 2/3$ at different displacement fields $D$. \textbf{c-d}, 2D-FFT of the oscillations. \textbf{e-f}, Diagonal conductance $G_\mathrm{d}$ oscillations as a function of TG voltage and dc bias. \textbf{g-h}, Transmitted resistance measured across the AD ($R_\mathrm{t}$ in dark red), longitudinal resistance measured outside the AD ($R_\mathrm{xx}$ in light red), diagonal conductance ($G_\mathrm{d}$ in dark blue) and Hall conductance ($G_\mathrm{xy}$ in light blue) as a function of TG voltage. The oscillations are taken along the dotted lines in their respective panels \textbf{a-c}.} 
 \label{fig:supp_6b}
 \end{center}
\end{figure*}

\begin{figure*}[htb!]
\renewcommand{\thefigure}{S16}
 \includegraphics[width = 0.95\textwidth]{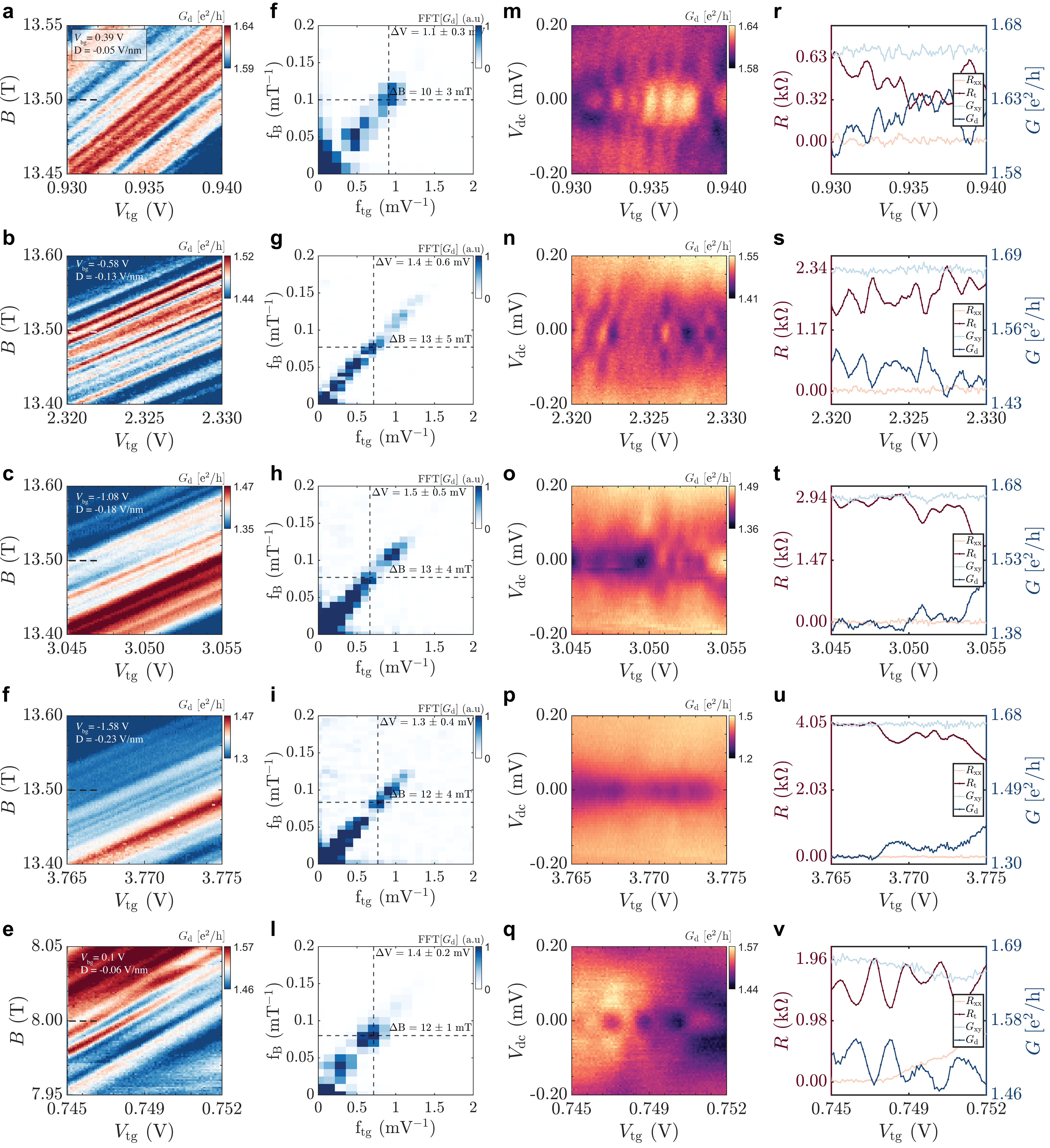}
 \begin{center}
 \caption{\textbf{Oscillations at $\nu = 5/3$.} \textbf{a-e}, Diagonal conductance $G_\mathrm{d}$ oscillations as a function of the TG voltage and magnetic field for filling factor $\nu = 5/3$ at different displacement fields $D$ and at different magnetic fields. \textbf{f-i}, 2D-FFT of the oscillations. \textbf{l-q}, Diagonal conductance $G_\mathrm{d}$ oscillations as a function of TG voltage and dc bias. \textbf{r-v}, Transmitted resistance measured across the AD ($R_\mathrm{t}$ in dark red), longitudinal resistance measured outside the AD ($R_\mathrm{xx}$ in light red), diagonal conductance ($G_\mathrm{d}$ in dark blue) and Hall conductance ($G_\mathrm{xy}$ in light blue) as a function of TG voltage. The oscillations are taken along the dotted lines in their respective panels \textbf{a-e}.} 
 \label{fig:supp_6d}
 \end{center}
\end{figure*}

\begin{figure*}[htb!]
\renewcommand{\thefigure}{S17}
 \includegraphics[width = 0.95\textwidth]{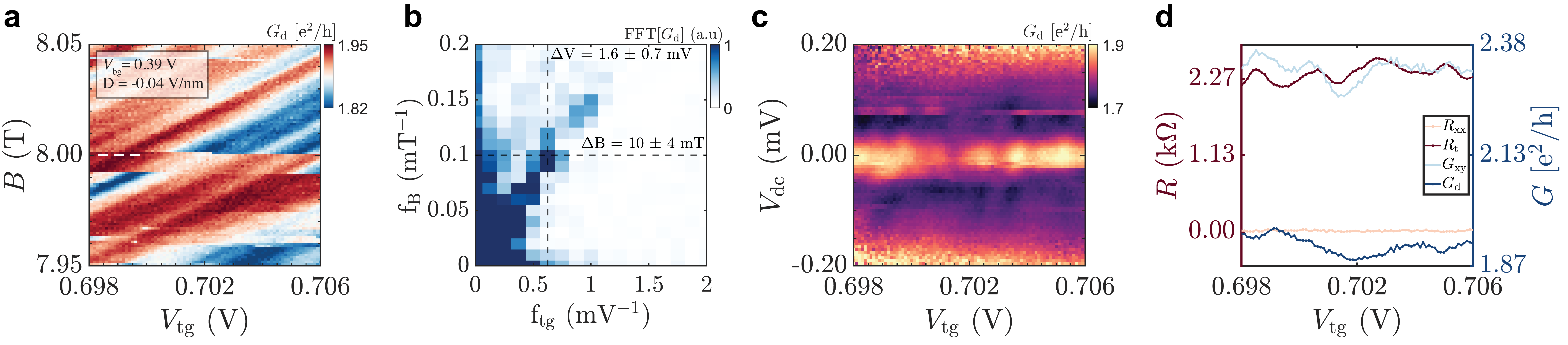}
 \begin{center}
 \caption{\textbf{Oscillations at $\nu = 7/3$.} \textbf{a}, Diagonal conductance $G_\mathrm{d}$ oscillations as a function of the TG voltage and magnetic field for filling factor $\nu = 7/3$ at different displacement fields $D$. \textbf{b}, 2D-FFT of the oscillations. \textbf{c}, Diagonal conductance $G_\mathrm{d}$ oscillations as a function of TG voltage and dc bias. \textbf{d}, Transmitted resistance measured across the AD ($R_\mathrm{t}$ in dark red), longitudinal resistance measured outside the AD ($R_\mathrm{xx}$ in light red), diagonal conductance ($G_\mathrm{d}$ in dark blue) and Hall conductance ($G_\mathrm{xy}$ in light blue) as a function of TG voltage. The oscillations are taken along the dotted lines in the respective panel \textbf{a}.} 
 \label{fig:supp_6e}
 \end{center}
\end{figure*}

\newpage
\clearpage
\newpage

\subsubsection*{Oscillations at $\nu = 8/3$}

\begin{figure*}[htb!]
\renewcommand{\thefigure}{S18}
 \includegraphics[width = 0.95\textwidth]{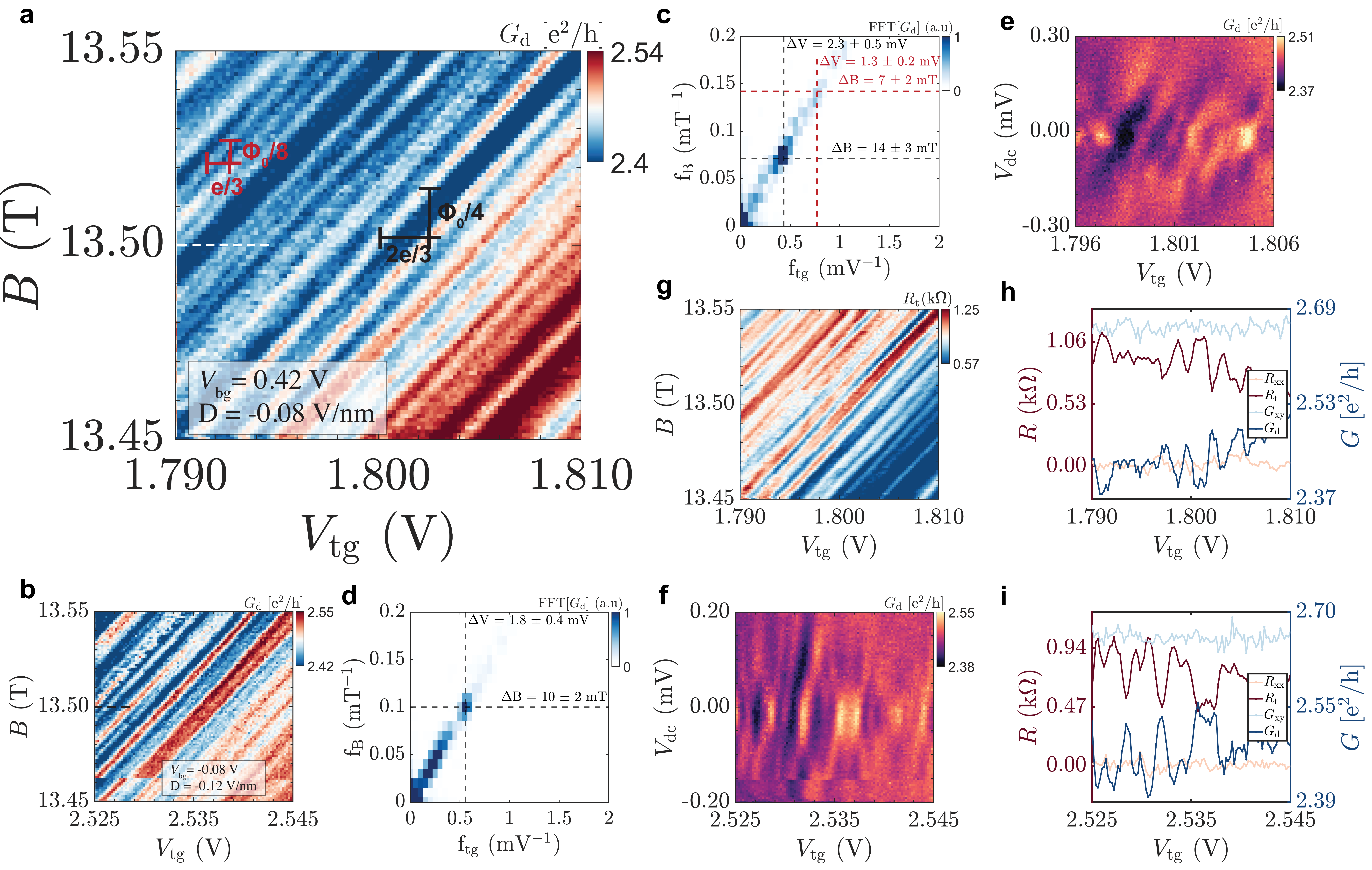}
 \begin{center}
 \caption{\textbf{Oscillations at $\nu = 8/3$.} \textbf{a-b}, Diagonal conductance $G_\mathrm{d}$ oscillations as a function of the TG voltage and magnetic field for filling factor $\nu = 8/3$ at different displacement fields $D$. The markers in panel \textbf{a} highlight the two different oscillation periods present at $\nu = 8/3$. \textbf{c-d}, 2D-FFT of the oscillations. The plot in panel \textbf{c} also reports the second peak in red. \textbf{e-f}, Diagonal conductance $G_\mathrm{d}$ oscillations as a function of TG voltage and dc bias. \textbf{g}, Transmitted resistance measured across the AD ($R_\mathrm{t}$ as a function of the TG voltage and magnetic field also displaying both oscillations. \textbf{h-i}, Transmitted resistance measured across the AD ($R_\mathrm{t}$ in dark red), longitudinal resistance measured outside the AD ($R_\mathrm{xx}$ in light red), diagonal conductance ($G_\mathrm{d}$ in dark blue) and Hall conductance ($G_\mathrm{xy}$ in light blue) as a function of TG voltage. The oscillations are taken along the dotted lines in their respective panels \textbf{a-b}.} 
 \label{fig:supp_6f}
 \end{center}
\end{figure*}

Of particular importance are the oscillations at filling factor $\nu = 8/3$, as they display two distinct types of periodicity. As highlighted in Fig. \ref{fig:supp_6f}a, the dominant oscillation corresponds to a quasiparticle tunneling charge of $q/e = 2/3$ and a flux periodicity of $\Delta\phi = \phi_0/4$, while in certain regions oscillations with smaller periodicity can be observed, corresponding instead to a quasiparticle tunneling charge of $q/e = 1/3$ and a flux periodicity of $\Delta\phi = \phi_0/8$. The 2D-FFT shows a main spectral peak corresponding to the larger-period ($2e/3$) oscillation, while a weaker secondary peak (highlighted in red) is associated with the $e/3$ oscillation. We interpret this as evidence that at $\nu = 8/3$ the fundamental quasiparticles carry charge $e/3$, and that this charge should, in principle, be observable. However, as discussed in the main text, the oscillation amplitude may depend on whether an even or odd number of quasiparticles is added to the AD, with one process dominating over the other. This mechanism can lead to the effective observation of a doubled charge, so that in most regimes only the stronger $2e/3$ oscillation is detected.

Finally, we note that the expected magnetic field oscillation period of the $e/3$ state, $\Delta B \sim 7 $ mT, is close to the minimum resolution achievable with our present magnet power supply. This instrumental limitation makes the direct detection of the smallest-period oscillation particularly challenging.

\newpage

\subsection{Renormalization Group Calculation}
\label{sub:RG}

In this section we provide a detailed calculation to supplement the discussion of tunnelling quasiparticle charges at $\nu=n+\frac23$. First, consider a single $\nu=\frac23$ edge which naively consists of a downstream unit-charge boson $\phi_1$ and an upstream fractional charge-$\frac13$ mode $\phi_{1/3}$. The quadratic action is described by
\begin{equation} \label{eq:quadratic_edge_lagrangian}
 \mathcal{L}=\frac{1}{4\pi}\left[-(\partial_t\phi_I)M_{IJ}\partial_x\phi_J + (\partial_x\phi_I)V_{IJ}(\partial_x\phi_J)\right]
\end{equation}
where $I,J\in\{1,1/3\}$, $M=\text{diag}(1,-3)$ and $V$ is a positive-definite $2\times2$ matrix. The insight of Kane, Fisher and Polchinski \cite{KaneFisherPolchinski} was to consider the charge-neutral operator $\mathcal{O}=e^{i[\phi_1+3\phi_{1/3}]}$ which tunnels an electron between the two edge modes. In particular, including this operator in $\mathcal{L}$ with a spatially-random prefactor, they showed that depending on the values of $V_{IJ}$ this perturbation may be relevant in the renormalization group (RG) sense and lead the system to a ``disorderd'' fixed point with a downstream charge-$\frac23$ mode and an upstream charge-neutral mode.

Now we consider the problem at $\nu=\frac53$ which has two downstream unit-charge modes $\phi_1,\phi_2$ (associated with fully filled lowest two LL's) and the upstream $\phi_{1/3}$. The starting point is again a Lagrangian of the form Eq.~\ref{eq:quadratic_edge_lagrangian}, but now with $I,J\in\{1,2,1/3\}$, $M=\text{diag}(1,1,-3)$ and $V$ a positive-definite $3\times3$ matrix. There are three important neutral tunneling operators to consider -- $\mathcal{O}_{1,2}=e^{i[\phi_{1,2}+3\phi_{1/3}]}$ which are the analogs of the $\mathcal{O}$ above, and $\tilde{\mathcal{O}}=e^{i[\phi_1-\phi_2]}$ which tunnnels electrons between the two integer modes.

The long-wavelength physics is determined by the operator with the lowest scaling dimension, which is in turn determined by $V_{IJ}$.
To build intuition, we first consider cases where $V_{12}=V_{1,1/3}=0$, so cases where $\phi_1$ has no density-density coupling to the other two modes. Following \cite{KaneFisherPolchinski}, we define $c=\frac{\sqrt{12}V_{2,1/3}}{3V_{22}+V_{1/3,1/3}}$, which must obey $|c|<1$ following from the positiveness of $V$. We find that in most of this range, for $c\in[-0.646,1]$, the most RG-relevant operator is $\tilde{\mathcal{O}}$. But for other values, $c\in[-1,-0.646]$, we have $\mathcal{O}_2$ as the most relevant. It is difficult to make further statements analytically, and without detailed microscopic knowledge of $V_{IJ}$, we turn to random sampling. We sample $V=X^TX$ with $X$ a $3\times3$ matrix of i.i.d. real Gaussian entires; we sample $V$ as a Wishart random matrix, which ensures positive-definiteness. 
The results show that in 75\% of the samples, $\tilde{\mathcal{O}}$ is the most RG-relevant operator, while $\mathcal{O}_1$ and $\mathcal{O}_2$ are each the most relevant in 6\% of the samples. In a further 5\% of the samples, the most relevant operator involves tunnelling more than one electron and the remaining 8\% of the samples have no relevant operators.

We conclude that very likely, $\tilde{\mathcal{O}}$ is the most relevant, so we consider where this leads the system. We start by focusing on $\phi_1$ and $\phi_2$ only, where a well-established procedure exists \cite{KaneFisher}. We write the $2\times2$ coupling of $\phi_{1,2}$ in $V$ as $V=v\mathbb{I}_2+\delta V$ with $\delta V$ traceless. Then, we fermionize these two edge modes to yield
\begin{equation} \label{eq:direct_fermionization}
 \mathcal{L}=i\psi_J^\dagger (\partial_t-v\partial_x)\psi_J + \xi(x)\psi_2^\dagger \psi_1 + \xi^*(x)\psi_1^\dagger\psi_2 + \delta V_{IJ} (\psi_I^\dagger\psi_I)(\psi_J^\dagger\psi_J)
\end{equation}
The advantage is that now the tunneling operator $\tilde{O}$ takes on the simple form $\psi_2^\dag \psi_1$ which we included above with a spatially random pre-factor $\xi(x)$. Finally, we write $\psi_a(x)=U_{ab}(x)\tilde{\psi}_b(x)$ with $U(x)$ a unitary chosen such that 
\begin{equation}
 \partial_x U(x) = \frac{i}{v}\begin{pmatrix}
 0 & \xi^*(x)\\
 \xi(x) & 0
 \end{pmatrix}U(x)
\end{equation}
Re-writing Eq.~\ref{eq:direct_fermionization} using this gives
\begin{equation}
 \mathcal{L}=i\tilde\psi^\dagger_J(\partial_t-v\partial_x)\tilde\psi_J + \mathcal{O}((\tilde\psi^\dagger\tilde\psi)^2)
\end{equation}
In this way, we have gauged away the tunnelling. The typical magnitude of $\xi(x)$ sets a lengthscale $l_\text{dis}$ beyond which the unitaries $U(x)$ are uncorrelated.
Re-writing the density operators in the new language gives
\begin{equation} \label{eq:density_with_rot)bosons}
 \begin{split}
 \rho_1=\frac{1}{2\pi}\partial_x\phi_1=\psi_1^\dagger\psi_1=U_{1a}(x)U_{1b}^*(x) \tilde\psi_b\tilde\psi_a=\frac{1}{4\pi}(\partial_x\tilde\phi_1+\partial_x\tilde\phi_2)+\\
 +R_1(x)\partial_x\tilde\phi_1+R_2(x)\partial_x\tilde\phi_2+R_3(x)e^{i[\tilde\phi_1-\tilde\phi_2]}+\text{h.c.}
 \end{split}
\end{equation}
Where $\tilde{\phi}_{1,2}$ are bosonizations of $\tilde{\psi}_{1,2}$ and all $R_k(x)$ are taken as spatially random with zero mean, correlated only on scales up to $l_\text{dis}$. The density $\rho_2$ obeys an analogous expression. After treating randomness, the expressions for physical densities in terms of the new bosons include purely random factors up to the constraint that the total charge density in $\tilde\phi_{1,2}$ is equal to that in $\phi_{1,2}$. Re-introducing the fractional mode, the edge is now described by (dropping the tilde on $\phi_{1,2}$)
\begin{equation} \label{eq:rotated_lagrangian_full}
 \begin{split}
 \mathcal{L}&=\mathcal{L}_0+\mathcal{L}_\text{random}\\
 \mathcal{L}_0&=\frac{1}{4\pi}\left[-(\partial_t\phi_I)M_{IJ}\partial_x\phi_J + (\partial_x\phi_I)V_{IJ}(\partial_x\phi_J)\right]\\
 V_{IJ}&=\begin{pmatrix}
 v & 0 & t\\
 0 & v & t\\
 t & t & u
 \end{pmatrix}\\
 \mathcal{L}_\text{random}&=R_1(x)\partial_x\phi_1\partial_x\phi_3+R_2(x)\partial_x\phi_2 \partial_x\phi_3+R_3(x)\partial_x\phi_3\left(e^{i[\phi_1-\phi_2]}+\text{h.c.}\right)\\&+R_4(x)\partial_x\phi_1\partial_x\phi_2 + R_5(x)e^{i[\phi_1+3\phi_3]}+\ldots
 \end{split}
\end{equation}
Here $M=\text{diag}(1,1,-3)$ and crucially Eq.~\ref{eq:density_with_rot)bosons} enforces this particular form of $V_{IJ}$. The upshot is that now all operators in $\mathcal{L}_\text{random}$ can be shown to have scaling dimension $\Delta>\frac32$, i.e. they are RG-irrelevant, regardless of the values in $V_{IJ}$. This means that there can be no relevant operator coupling the fractional mode to the integer edge. The strong hybridization between integer modes blocks the Kane-Fisher-Polchinski mechanism and leaves an upstream charge-$\frac13$ mode intact.

By analogous reasoning, one may consider $\nu=n+\frac23$ for $n>0$ integer. Then we have $n+1$ integer modes $\phi_1,\ldots,\phi_{n+1}$ and if we assume they can all couple strongly, the integer modes will flow into a disordered fixed point while the fractional mode would remain decoupled. But as noted above, some of these couplings may be very weak in practice and thus for some $n$, this physics could only be visible at temperatures well below what is used in experiments. Alternatively, if the tunneling is weak, the length $l_\text{dis}$ over which $R_k(x)$ are uncorrelated becomes large, and the terms in $\mathcal{L}_\text{random}$ no longer have random pre-factors which means they can become relevant if they have $\Delta\leq2$, which is a weaker requirement than the $\Delta\leq\frac32$ needed for an operator with a random pre-factor to be relevant.

Note that despite the lack of RG-relevant couplings, we can still expect $\phi_{1/3}$ to equilibrate with the integer edges in experiments at finite temperature (to still see quantized edge transport), but we should expect the equilibration length to diverge at low temperatures \cite{Ma_2020,KaneFisher}. So at sufficiently low temperatures, the quantization is expected to be broken, but the exact temperature where this happens depends on microscopic details, the calculation of which we do not attempt.

\end{document}